\documentclass[aps,prl,reprint,showpacs,superscriptaddress]{revtex4-1}

\usepackage{graphicx}
\usepackage{color}
\usepackage[colorlinks, linkcolor=blue, anchorcolor=blue, citecolor=blue, urlcolor=blue]{hyperref}
\usepackage{multirow}
\usepackage{tabularx}
\usepackage{rotating}
\usepackage{threeparttable}
\usepackage{longtable}

\newcommand{\cmao}{CeMgAl$_{11}$O$_{19}$}
\newcommand{\lmao}{LaMgAl$_{11}$O$_{19}$}
\newcommand{\pmao}{PrMgAl$_{11}$O$_{19}$}

\begin{document}


\title{U(1) Dirac quantum spin liquid candidate in triangular-lattice antiferromagnet CeMgAl$_{11}$O$_{19}$}



\author{Yantao Cao}
\affiliation{Songshan Lake Materials Laboratory, Dongguan 523808, China}
\affiliation{Institute of Physics, Chinese Academy of Sciences, Beijing 100190, China}

\author{Akihiro Koda}
\affiliation{Muon Science Laboratory, Institute of Materials Structure Science, KEK, Tokai, Ibaraki 319-1106, Japan}

\author{M. D. Le}
\affiliation{ISIS Neutron and Muon Source, Rutherford Appleton Laboratory, Chilton, Didcot OX11 0QX, United Kingdom}

\author{V. Pomjakushin}
\affiliation{Laboratory for Neutron Scattering and Imaging LNS, Paul Scherrer Institute, Villigen CH-5232, Switzerland}

\author{Benqiong Liu}
\address{Institute of Nuclear Physics and Chemistry, China Academy of Engineering Physics (CAEP), Mianyang 621999, China}

\author{Zhendong Fu}
\affiliation{Songshan Lake Materials Laboratory, Dongguan 523808, China}

\author{Zhiwei Li}
\affiliation{School of Physical Science and Technology, Lanzhou University, Lanzhou 730000, China}

\author{Jinkui Zhao}
\email{jkzhao@sslab.org.cn}
\affiliation{School of Physical Sciences, Great Bay University, Dongguan 523808, China}
\affiliation{Songshan Lake Materials Laboratory, Dongguan 523808, China}
\affiliation{Institute of Physics, Chinese Academy of Sciences, Beijing 100190, China}

\author{Zhaoming Tian}
\email{tianzhaoming@hust.edu.cn}
\affiliation{School of Physics and Wuhan National High Magnetic Field Center,
Huazhong University of Science and Technology, Wuhan 430074, China}

\author{Hanjie Guo}
\email{hjguo@sslab.org.cn}
\affiliation{Songshan Lake Materials Laboratory, Dongguan 523808, China}



\date{\today}

\begin{abstract}
  Quantum spin liquid represents an intriguing state where electron spins are highly entangled yet spin fluctuation persists even at 0 K. Recently, the hexaaluminates \textit{R}MgAl$_{11}$O$_{19}$ (\textit{R} = rare earth) have been proposed to be a platform for realizing the quantum spin liquid state with dominant Ising anisotropic correlations. Here, we report detailed low-temperature magnetic susceptibility, muon spin relaxation, and thermodynamic studies on the CeMgAl$_{11}$O$_{19}$ single crystal. Ising anisotropy is revealed by magnetic susceptibility measurements. Muon spin relaxation and ac susceptibility measurements rule out any long-range magnetic ordering or spin freezing down to 50 mK despite the onset of spin correlations below $\sim$0.8 K. Instead, the spins keep fluctuating at a rate of 1.0(2) MHz at 50 mK. Specific heat results indicate a gapless excitation with a power-law dependence on temperature, $C_m(T) \propto T^{\alpha}$. The quasi-quadratic temperature dependence with $\alpha$ = 2.28(4) in zero field and linear temperature dependence in 0.25 T support the possible realization of the U(1) Dirac quantum spin liquid state.
\end{abstract}


\maketitle


\section{Introduction}\label{sec:3}
The triangular-lattice antiferromagnet (TLAF) is a fertile playground for searching exotic quantum phases and excitations, such as the spin supersolid state \cite{Heidarian2010,Gao2022,Xiang2024} and quantum spin liquid (QSL) state \cite{Balents2010,Savary2016,Broholm2020}. In a QSL, the spins resist ordering even at zero kelvin due to strong frustration and quantum fluctuations. Instead, they form a highly entangled state that exhibits fractional excitations as opposed to the traditional magnons. Despite being heavily investigated since the early work of Anderson \cite{Anderson1973,Anderson1987}, there is no consensus on whether any real material has realized this intriguing state. In fact, many TLAFs show a magnetically ordered state as the temperature is lowered towards 0 K \cite{Mekata1993,Shirata2012}. Even for a spin-disordered state, the presence of antisite disorders may complicate the interpretation of experimental results \cite{Kimchi2018,Kimchi2018-2}. One prominent example is the compound YbMgGaO$_4$ whose nonmagnetic ions Mg and Ga are randomly distributed and can subsequently influence the Yb-O-Yb bond, resulting in QSL-like behavior \cite{Ma2018,Li2019,Li2019-2}.

Various kinds of QSLs have been proposed, which can primarily be classified as either gapped or gapless \cite{Wen2002}. For the triangular lattice system, the one characterized by gapless emergent fermionic spinons forming a Fermi surface \cite{Nagaosa1990} or Dirac cone \cite{Ran2007} has been proposed. Several compounds have been suggested to realize these intriguing states \cite{Dai2021,Bag2024}. It was recently realized that the layered hexaaluminate structure with the general formula of \textit{R}MgAl$_{11}$O$_{19}$ (space group $P6_3/mmc$), where the magnetic rare earth ions \textit{R} decorate a triangular sublattice, could also host a QSL state \cite{Ashtar2019,Bu2022,Cao2024,Ma2024}. Magnetic susceptibility and thermodynamic studies on  polycrystalline PrZnAl$_{11}$O$_{19}$ showed no magnetic ordering, nor any spin freezing down to 50 mK, despite a large spin-spin interaction. The quasi-quadratic power-law dependence of the magnetic specific heat on temperature, along with a broad, continuum-like low energy excitation as revealed by inelastic neutron scattering measurement are in line with a U(1) Dirac QSL state \cite{Bu2022}.
Further studies on single crystals of the isostructural compound PrMgAl$_{11}$O$_{19}$ unveil a pronounced Ising anisotropy with moments lying along the crystallographic \textit{c} axis \cite{Cao2024}.
Pr$^{3+}$ is a non-Kramers ion so that the ground state is not protected by the time reversal symmetry. Indeed, high-field electron spin resonance (ESR) measurement on PrMgAl$_{11}$O$_{19}$ revealed a zero-field gap of $\sim$0.1 meV between the two low-lying singlets. Thus, the longitudinal spin component behaves as a dipole while the transverse components behave as multipoles \cite{Shen2019,chen2019}.

So far, no detailed investigation on the magnetic properties of Kramers ion in this series of compounds has been reported. Here, we present comprehensive thermodynamic, magnetic susceptibility and muon spin relaxation ($\mu$SR) measurements on CeMgAl$_{11}$O$_{19}$ based on Ce$^{3+}$ with an effective spin $J_\mathrm{eff}$ = 1/2. This compound shows a marked Ising anisotropy. Antiferromagnetic correlations develop below $\sim$0.8 K, whereas the spins do not order nor freeze down to 50 mK, but fluctuate at a rate of 1.0(2) MHz, reflecting a highly entangled state. Gapless low energy excitations are revealed by specific heat measurements, which show a quasi-quadratic temperature dependence at zero field and a linear temperature dependence in a magnetic field of 0.25 T, in line with a U(1) Dirac QSL.

\section{Materials and method}\label{sec:4}
Centimeter-sized single crystals of CeMgAl$_{11}$O$_{19}$ were grown following the same procedure as that described for PrMgAl$_{11}$O$_{19}$ \cite{Cao2024}. The obtained single crystals were postannealed in a flowing O$_2$ atmosphere at 1000 $^{\circ}$C for 24 hours to avoid any possible oxygen deficiency and show the robustness of the Ce$^{3+}$ state in our sample; see the Supplementary Note 2 for more details.
Single crystal X-ray diffraction (XRD) measurements were performed on an XtaLAB Synergy diffractometer (Rigaku) at room temperature using the Mo-$K_\alpha$ radiation. The experimental conditions are tabulated in the Supplementary Tab. S1. Neutron powder diffraction measurements were performed on the HRPT diffractometer at the Paul Scherrer Institut (PSI), Switzerland. Inelastic neutron scattering on powders was performed with different energies on the SEQUOIA and MARI instruments. JANA \cite{Jana} and FULLPROF \cite{Fullprof} softwares were used for crystal structure refinements.

DC magnetic susceptibility between 2 and 350 K was measured using the vibrating sample magnetometer (VSM) option of the Physical Property Measurement System (PPMS, Quantum Design). AC magnetic susceptibility
between 0.05 and 15 K was measured using the ACMS-II and ACDR options of the PPMS equipped with a dilution insert. A driven field of 1-3 Oe in amplitude was used. Heat capacity measurements were carried out on the PPMS using the relaxation method.

Muon spin relaxation measurements were performed on the D1 spectrometer at J-PARC, Japan. The powders were mixed with the GE-varnish and attached to a silver plate in order to have good thermal contacts at 50 mK. One advantage of using the polycrystal rather than single crystal is that we avoid the risk of probing the direction without appreciable dynamics, i.e., the internal fields are along the initial muon spin direction. Moreover, our specific heat measurements on the single crystal and polycrystal indicate that there is no noticeable difference in low energy excitations; see the Supplementary Note 5. The experimental asymmetry, which is proportional to the muon spin polarization, is defined as $A(t) = \frac{F(t) - \alpha B(t)}{F(t) + \alpha B(t)}$, where $F(t)$ ($B(t)$) is the number of positrons arriving at the forward (backward) detector at time \textit{t}. The parameter $\alpha$ reflects the different counting efficiencies for the forward and backward detectors. For longitudinal-field measurements, a longitudinal field was applied along the initial muon spin direction.

\section{Results}

Following the same refinement procedure as for the isostructural compound PrMgAl$_{11}$O$_{19}$ \cite{Cao2024}, we found that a small amount of Ce ions ($\sim$13\%) are displaced from the 2\textit{d} site towards the 6\textit{h} site. There is also a mixing of Al and Mg at the 4\textit{f} site. The refined structure parameters are found in the Supplementary Tab. S1 and Fig. S2. The presence of disorder at the Ce site and the 4\textit{f} site may seem quite unfavorable for a quantum spin liquid state and thus complicate our analyses. However, detailed considerations suggest that their influence on the spin dynamics may be negligible, as will be discussed later.

\begin{figure}[tbh]
  \centering
  \includegraphics[width=1\columnwidth]{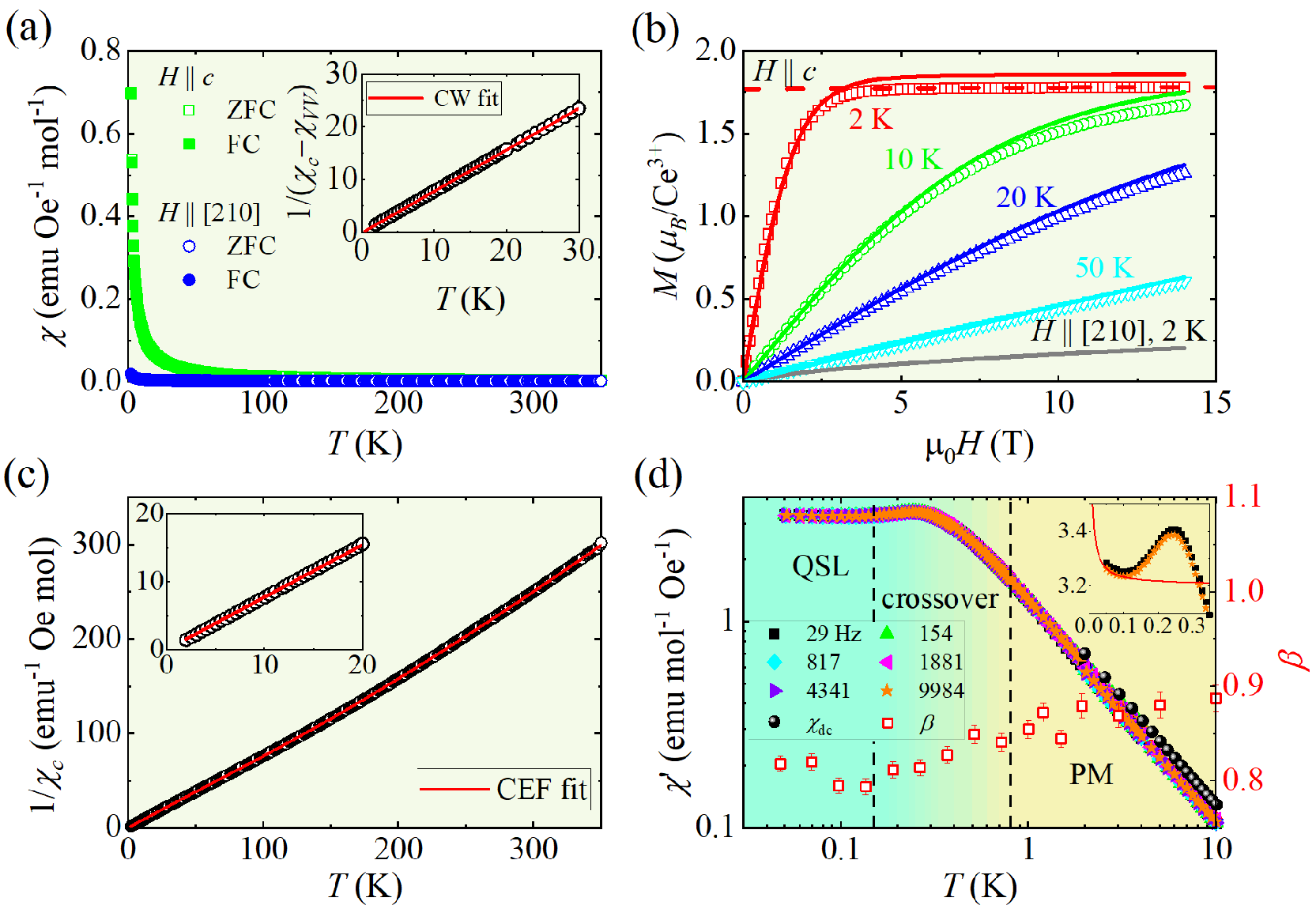}\\
  \caption{\textbf{Magnetic susceptibility and isothermal magnetization measurements.} (a) Temperature dependence of the DC magnetic susceptibility along different directions. The inset shows a low-temperature Curie-Weiss fit. (b) Isothermal magnetization measured at various temperatures. The field was applied along the \textit{c} axis. The magnetization along the [210] direction at 2 K is shown for comparison. (c) Temperature dependence of the inverse susceptibility along the \textit{c} direction. The inset highlights the low-\textit{T} region. The solid lines in (b,c) are fits according to the CEF model. The dashed line in (b) is a linear fit to extract the saturation magnetization and Van Vleck paramagnetic susceptibility. (d) Temperature dependence of the ac susceptibility measured at various frequencies. The inset highlights the low-\textit{T} region with a modified Curie fit. The exponent from the $\mu$SR fit is also shown. The tentative phase diagram shows a QSL region where the susceptibility is temperature independent, a paramagnetic (PM) region where the susceptibility follows a Curie behavior, $\propto$ 1/\textit{T}, and a crossover region in between.}\label{sus}
\end{figure}

Figure \ref{sus}(a) shows the temperature dependence of the magnetic susceptibility with magnetic field applied along different crystallographic directions. No difference or bifurcation is observed for the zero-field-cooled (ZFC) and field-cooled (FC) curves down to 2 K, consistent with a paramagnetic state. Moreover, a pronounced anisotropy is evident at low temperatures, which is more obvious from the isothermal magnetization measurements as shown in Fig. \ref{sus}(b). At 2 K and 14 T, the magnetization along the \textit{c} direction is about 9 times larger than that along the [210] - or \textit{a}* in the reciprocal space - direction.

For rare earth ions, their magnetic behaviors are strongly influenced by the surrounding crystal electric field (CEF). Specifically, for the current system, the $D_{3h}$ symmetry at the 2\textit{d} site will split the lowest multiplet of Ce$^{3+}$, $^{2}F_{5/2}$, into three Kramers doublets. The crystal field Hamiltonian can be expressed as $\mathcal{H}_{CEF} = \sum_{l,m} B_l^mO_l^m$, where $O_l^m$ are the Stevens operators \cite{Stevens1952,Hutchings1964}, and $B_l^m$ are parameters that can be determined experimentally. Note that X-ray photoelectron spectroscopy (XPS) measurements on the oxygen-annealed sample indicate a Ce$^{3+}$ valence state in CeMgAl$_{11}$O$_{19}$, see the Supplementary Note 2. For those Ce$^{3+}$ ions at the 2\textit{d} site, only $B_2^0$ and $B_4^0$ are nonzero, which will result in three doublets consisting of pure $|\pm 5/2\rangle$, $|\pm 3/2\rangle$, and $|\pm 1/2\rangle$ in the $|m_J\rangle$ representation. The $|\pm 5/2\rangle$ state is the ground state due to the out-of-plane Ising anisotropy.
The saturation moment for this ground state amounts to 2.14 $\mu_B$/Ce, which is much larger than the experimental value (1.77 $\mu_B$) at 2 K after subtracting the Van Vleck contribution; see Fig. \ref{sus}(b). Taking the 13\% displaced Ce$^{3+}$ ions into account, and assuming an extreme case of a pure $J_z = \pm$ 1/2 state, the calculated saturation magnetization of 1.92 $\mu_B$ is still larger than the experimental one; see the Supplementary Note 3 for more details. Note that this is based on a weak coupling scheme (Russell-Saunders scheme) in the $|J, m_J\rangle$ basis. The failure of the weak coupling scheme is verified by a fit to the inverse susceptibililty $\chi_c^{-1}$ using the PyCrystalField package \cite{Scheie}; see the Supplementary Fig. S5.

Alternatively, the experimental data can be well described by a model based on the intermediate coupling scheme using the $|L,S,m_L,m_S\rangle$ basis. The calculated susceptibility is corrected by taking into account the interactions with neighboring ions such that $\chi_\mathrm{cal}^c = \chi_\mathrm{CEF}^c/(1-\varepsilon\chi_\mathrm{CEF}^c)$. As shown in Fig. \ref{sus}(c) and the inset, the simulated curve agrees well with the data across the whole temperature range. A negative $\varepsilon$ of -0.013 T/$\mu_B$ indicates an overall antiferromagnetic interaction. Alternative, it may also be due to the omission of the randomly displaced Ce ions; see the Supplementary Note 4 for more discussions. The crystal field parameters and corresponding eigenvectors are found in the Supplementary Tab. S6.
The first excited state is 36.2 meV above the ground state, so that the low temperature properties can be described by an effective spin $J_\mathrm{eff}$ = 1/2 state. Therefore, we analyze the low temperature susceptibility ( \textit{T} $\le$ 30 K) by a Curie-Weiss fit such that $(\chi_c - \chi_{VV})^{-1} = (T - \theta_{CW})/C$, where $\chi_{VV}$ = 5.26 $\times 10^{-4}$ emu Oe$^{-1}$ mol$^{-1}$ is the Van Vleck paramagnetic susceptibility deduced from the MH curve at 2 K. The fit yields an effective moment of $\sqrt{8C}$ = 3.19 $\mu_B$/Ce. From $\mu^c_{eff} = g_c\sqrt{S(S+1)}$ and \textit{S} = 1/2, one obtains $g_c$ = 3.68, which is very close to the $g_c$ value extracted from the CEF ground state ($g_c$ = 3.71 and $g_{ab}$ = 0.39). The extracted effective moment is larger than that expected for a free ion because of the large contributions from the $|m_L = \pm3, m_S = \mp1/2\rangle$ state in the ground state; see the Supplementary Tab. S6.
The fit further yields a CW temperature of 0.17 K.
A positive $\theta_{CW}$ seems unusual since there will be no frustration and an ordered state is expected. Note that the positive $\theta_{CW}$ along \textit{c}-axis should be intrinsic as it is reproducible in several measurements. The origin should be due to the superexchange interaction via the intermediate oxygen ion. As a comparison, the strength of the dipole interaction can be estimated as $D = \mu_{sat}^2/r_{nn}^3 \sim$ 0.01 K, where $\mu_{sat}$ is the saturation moment, and $r_{nn}$ is the nearest neighbor distance between the Ce ions. A similar positive $\theta_{CW}$ was also observed in the Kagome QSL candidate Ca$_{10}$Cr$_{7}$O$_{28}$, suggesting a complex frustration in the sample \cite{Balz2016}. It is possible that there exist antiferromagnetic couplings within the \textit{ab}-plane, as was observed from the inelastic neutron scattering measurements in the spin polarized state \cite{Gao2024}, which leads to frustration and prevents the system from ordering. Note that the Ising character with very small susceptibility within the \textit{ab}-plane renders a reliable CW fit impossible.

Based on the crystal field parameters and $\varepsilon$, the magnetizations at various temperatures can be calculated. As shown in Fig. \ref{sus}(b), the calculated curves agree well with the data, indicating that the CEF model captures at least the ground state of the CEF scheme. It is worth noting that we also try to resolve the CEF excitations using more direct measurements such as inelastic neutron scattering, which turns out to be very challenging due to the low concentration of Ce$^{3+}$ ions. As shown in the Supplementary Fig. S7, no discernible CEF excitations are observed up to 60 meV. In fact, higher $E_i$ up to 1 eV was used but yielded no positive results.

In order to probe the low temperature magnetic properties, ac susceptibility was measured down to 50 mK. As shown in Fig. \ref{sus}(c), the real component, $\chi'$, increases with decreasing temperature and shows a broad hump at $\sim$0.25 K. At lower temperatures, it becomes almost temperature independent. The slight upturn below 0.1 K may be due to a tiny amount of paramagnetic impurities (uncorrelated or orphan Ce$^{3+}$ spins). A modified Curie fit, $\chi(T) = p\cdot C/T + \chi_0$, to the data below 0.1 K yields a small concentration (\textit{p}) of impurities, about 0.2\%. Here, \textit{C} is the Curie constant extracted from the fit above 2 K, and $\chi_0$ represents a temperature independent term. The appearance of the hump indicates the development of antiferromagnetic correlations. More importantly, the absence of any frequency dependence rules out the possibility of spin glass transition. The nonzero susceptibility at low temperatures is consistent with a gapless excitation as observed in many QSL candidates \cite{Li2016,Isono2014}.

\begin{figure}
  \centering
  \includegraphics[width=1\columnwidth]{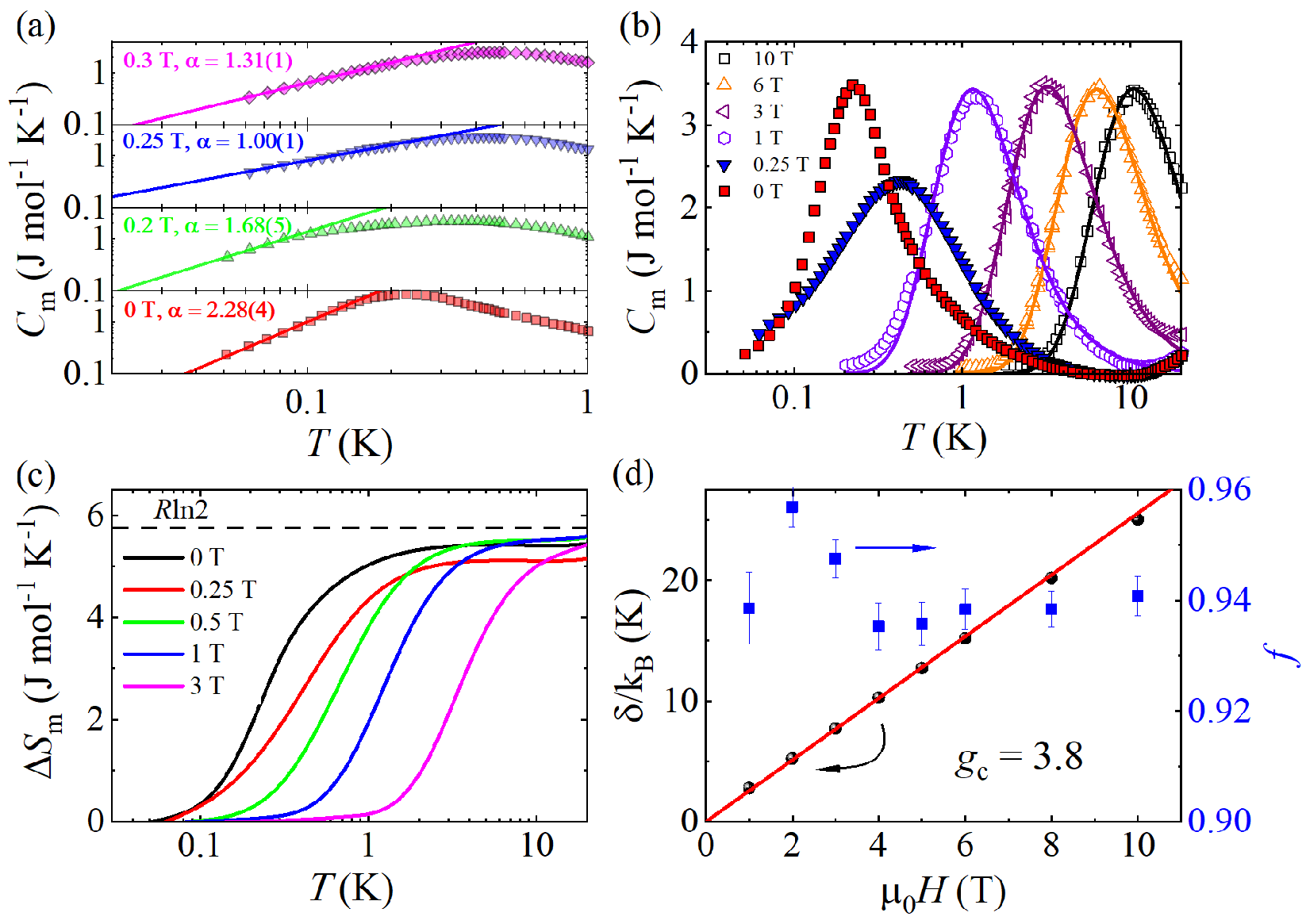}\\
  \caption{\textbf{Heat capacity measurements.} (a) Temperature dependence of the specific heat measured at various magnetic fields applied along the \textit{c}-axis. The phonon contribution obtained from the nonmagnetic LaMgAl$_{11}$O$_{19}$ has been subtracted. (b) High field results together with Schottky fits. Zero-field and 0.25-T data are also shown for comparison. (c) Magnetic entropy change obtained by integrating $C_m/T$ over \textit{T} at various magnetic fields. (d) Magnetic field dependence of the energy gap obtained from the Schottky fit in (b).}\label{HC}
\end{figure}

The low-energy excitations were further probed by specific heat measurements. One advantage of the studied compound is that there is no upturn of the specific heat at low temperatures due to the Schottky anomaly from the nuclear moments. Therefore, the magnetic contribution can be obtained by subtracting the phonon contributions using LaMgAl$_{11}$O$_{19}$ as the reference sample. The temperature dependence of the magnetic specific heat, $C_m$, was obtained with magnetic field applied along the \textit{c} axis and is shown in Fig. \ref{HC}(a). The temperature dependence of the change of the magnetic entropy, $\Delta S_m(T)$, is shown in Fig. \ref{HC}(c). It reaches 94\% of \textit{R}ln2 at 5 K, indicating that there is no appreciable entropy below 50 mK, thus ruling out possible long-range magnetic transitions at lower temperatures. At 0.25 T, the saturated $\Delta S_m$ is 89\% of \textit{R}ln2, suggesting that more entropies are retained at low temperatures.

In zero field, no sharp, $\lambda$-shaped peak associated with long-range magnetic ordering is observed down to 50 mK. The broad peak at $\sim$0.25 K is suggestive of short range correlations. Below 0.25 K, $C_m$ exhibits a power-law dependence on the temperature, $C_m = AT^{\alpha}$, with $\alpha$ = 2.28. First, we note that several ordered system can show a power-law dependence stemmed from the spin wave dispersions, such as a $T^3$ dependence for the gapless antiferromagnet and a $T^{3/2}$ dependence for a ferromagnet \cite{Gopal}. For a QSL, a linear \textit{T} or $T^{2/3}$ dependence for a QSL with spinon Fermi surface \cite{Motrunich2005,Yamashita2008}, and $T^{3}$ dependence for a Coulombic QSL \cite{Savary2012} are proposed. However, none of these is consistent with the observed quasi-quadratic temperature dependence for CeMgAl$_{11}$O$_{19}$. To the best of our knowledge, only a U(1) Dirac QSL can result in such a $T^2$ dependence due to the Dirac nodes \cite{Ran2007}.
Moreover, theory predicts a linear temperature dependence when $k_B T \ll \mu_B H$ due to the formation of Fermi pocket \cite{Ran2007}. As shown in Fig. \ref{HC}(a), the exponent $\alpha$ decreases with increasing fields, and reaches a value of 1 at 0.25 T ($\mu_B H/k_B$ = 0.17 K) below 0.2 K.
In addition, the ratio of the coefficients for the \textit{T} term in fields and $T^2$ term in zero field is expected to be 0.21\textit{H} = 0.053 at \textit{H} = 0.25~T \cite{Ran2007}. From the fits in Fig. \ref{HC}(a), the ratio amounts to 0.041, agreeing reasonably well with the prediction.
At higher fields, the moments are fully polarized, resulting in a Schottky behavior. Fits of the two-level Schottky model, $C_m = f\cdot R(\frac{\delta}{k_BT})^2\frac{\mathrm{exp}(\delta/k_BT)}{[1+\mathrm{exp}(\delta/k_BT)]^2}$, to the data at different fields are shown in Fig. \ref{HC}(b), where \textit{R} is the ideal gas constant, $k_B$ is the Boltzmann's constant, $\delta$ is the opening gap, and \textit{f} is the fraction of free ions. Using $\delta = g_c\mu_B\mu_0H$, the extracted $g_c$ = 3.8 is consistent with the magnetic susceptibility analyses.

\begin{figure}
  \centering
  \includegraphics[width=1\columnwidth]{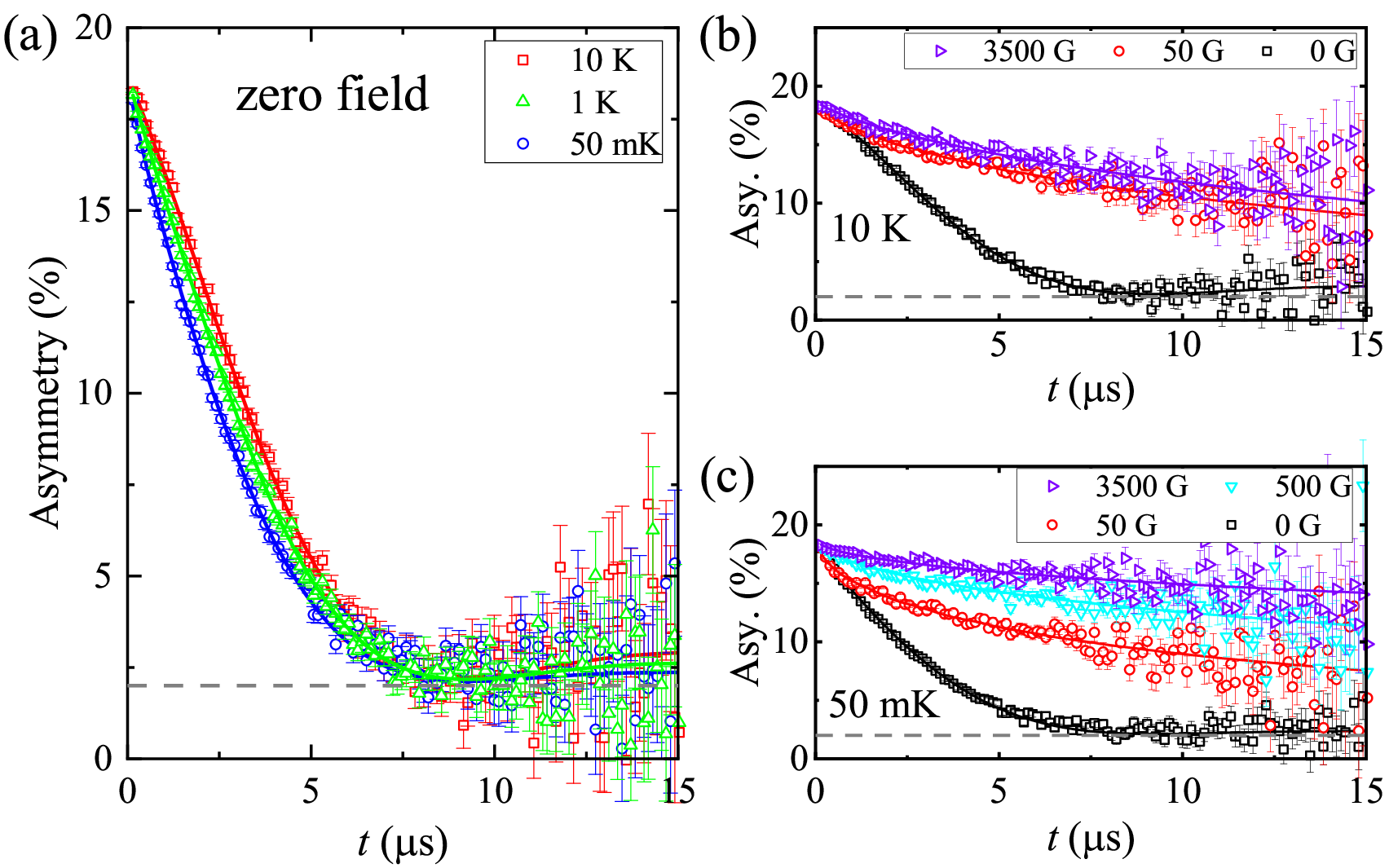}\\
  \caption{\textbf{Zero-field and longitudinal-field $\mu$SR spectra.} (a) Typical ZF time spectra for CeMgAl$_{11}$O$_{19}$. (b) and (c) LF spectra measured at 10 and 0.05 K, respectively. The solid lines are the fits; see the text for details. The dashed lines indicate the background position, $A_{bg}$.}\label{spec}
\end{figure}

Figure \ref{spec}(a) shows the $\mu$SR time spectra measured in zero field (ZF). At 10 K, the spectrum exhibits a Kubo-Toyabe-like behavior with a dip at around 8 $\mu$s. With decreasing temperatures, the initial relaxation becomes faster, but the overall spectrum shape remains Kubo-Toyabe like. No spontaneous muon spin precession was observed, thus ruling out the formation of any long-range magnetically ordered state. The dynamic nature is further corroborated by longitudinal field (LF) measurements. As shown in Fig. \ref{spec}(b), the asymmetry is largely recovered after $\sim$1 $\mu$s and slowly relaxed in a small LF of 50 G at 10 K. This slow relaxation is almost independent of the LF, indicating a fast fluctuation of the electronic spins at high temperatures. At the base temperature of 50 mK, the asymmetry is gradually recovered with increasing fields, but a clear relaxation can still be observed at 3500 G.

\begin{figure}
  \centering
  \includegraphics[width=1\columnwidth]{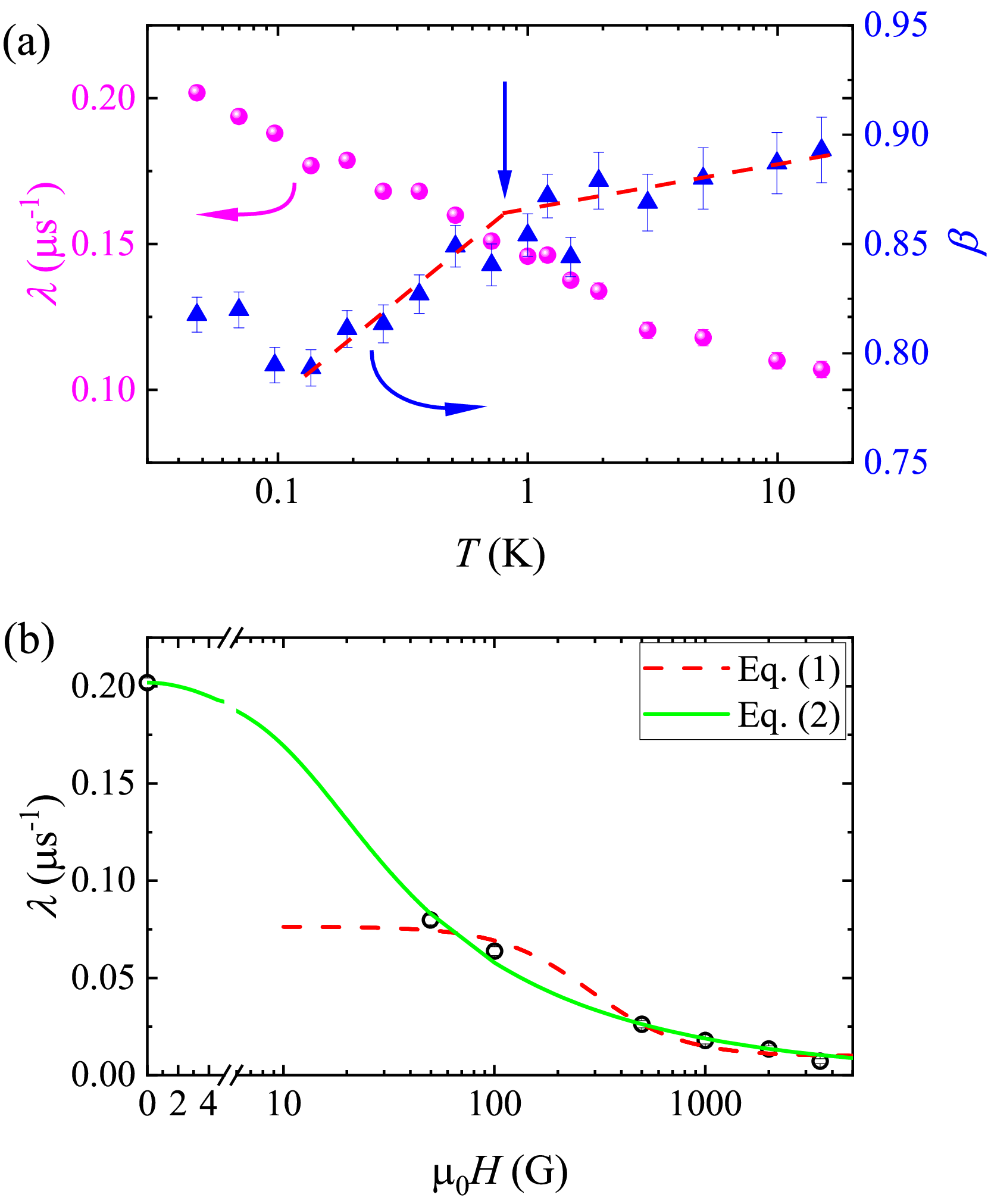}\\
  \caption{\textbf{$\mu$SR fitting parameters.} (a) Temperature dependence of the muon spin relaxation rate $\lambda$ and exponent $\beta$ in zero field. The vertical arrow indicates a kink around 0.8 K for $\beta$. The red dashed lines are guide to the eyes.(b) External field dependence of the muon spin relaxation rate. The red dashed curve and green solid curve are fits according to Eq. (1) and Eq. (2), respectively.}\label{para}
\end{figure}

To have a quantitative understanding of the dynamics, the ZF time spectra were analyzed with the function
$$A(t) = A_1G_{KT}(t,\Delta)\mathrm{exp}[-(\lambda t)^\beta] + A_{bg},$$
where $G_{KT}(t,\Delta)$ is the Kubo-Toyabe function with a Gaussian broadening width $\Delta$ and originates mostly from the nuclear moments of Al (\textit{I} = 5/2) \cite{Guo2013}. The stretched exponential term describes the relaxation channel from the electronic spins. $A_1$ and $A_{bg}$ represent the fraction of muons stopped in the sample and the Ag holder, respectively. The initial asymmetry $A(0)$, $A_{bg}$ and $\Delta$ were fixed to the values obtained at 10 K. The $\Delta$ amounts to 0.19 $\mu s^{-1}$, corresponding to a distribution width of $\Delta/\gamma_{\mu}$ = 2.23 G for the static internal fields. Here, $\gamma_{\mu}/2\pi$ = 135.5 MHz T$^{-1}$ is the gyromagnetic ratio of muon.
The temperature dependence of the relaxation rate $\lambda$ and exponent $\beta$ are shown in Fig. \ref{para}(a).
The relaxation rate increases monotonically with decreasing temperatures, indicating a continuous slowing down of the Ce$^{3+}$ spins. Usually, one would expect a plateau in the temperature dependence of $\lambda$ in the frustrated systems due to the persistent fluctuations. And since such a plateau is observed in the ac susceptibility, the onset temperature is expected to be higher in the $\mu$SR result. This discrepancy may be attributed to the different sample forms used for the experiments, i.e., a single crystal and a polycrystal for the ac susceptibility and $\mu$SR measurements, respectively. When a polycrystal was used for the ac susceptibility measurement (data not shown), the plateau is smeared out, and a continuous increase is observed at lower temperatures. Future $\mu$SR experiment on single crystals will be needed to clarify this point.

The exponent $\beta$ is below 1 in the measured temperature range, indicating a distribution of the relaxation time. $\beta$ is almost independent of temperature above 1 K, but exhibits a kink at about 0.8 K, suggesting an onset of spin correlations \cite{Ishant2024}.  Note that $\beta$ is much larger than 1/3, which was sometimes misinterpreted as evidence for the absence of a spin glass state from the viewpoint of $\mu$SR. For a canonical spin glass, the 1/3 value is observed at the freezing temperature. $\beta$ could be larger at higher and lower temperatures \cite{Campbell1994,Keren1996}. For more complicated systems such as YbMgGaO$_4$, $\beta$ is also larger than 1/3 \cite{Li2016}, but ac susceptibility shows a clear frequency dependence. In our case, however, a combination with the ac susceptibility results rules out a spin glass ground state. The temperature dependence of $\beta$ is plotted together with the susceptibility in Fig. \ref{sus}(d), along with a tentative phase diagram.

In a longitudinal field larger than 50 G, which is about 25 times larger than $\Delta/\gamma_\mu$, the $G_{KT}(t)$ term will be nearly flat and equals to 1, thus, the LF spectra were fitted using $A(t) = A_1\mathrm{exp}[(-\lambda t)^{\beta}] + A_{bg}$. The magnetic field dependence of the relaxation rate is shown in Fig. \ref{para}(b). A fit of the modified Redfield formula \cite{Guo2014}
\begin{equation}\label{}
  \lambda(H) = \frac{2\Delta^2\nu}{\nu^2+(\gamma_\mu \mu_0H)^2} + \lambda_0
\end{equation}
does not satisfactorily describe the data, suggesting that the system is not in the motional narrowing regime of $\nu/\Delta \gg 1$, where $\nu$ is the fluctuating rate of Ce$^{3+}$ spins. The relaxation rate can be expressed more generally as

\begin{equation}\label{}
  \lambda(H) = 2\Delta^2\tau^x\int_0^\infty t^{-x}\mathrm{exp}(-\nu t)\mathrm{cos}(\gamma_\mu \mu_0Ht)dt,
\end{equation}
where $\tau$ is an early time cutoff \cite{Keren2001,Li2016}. The best-fit yields \textit{x} = 0.53(3), $\Delta$ = 9(1) MHz and $\nu$ = 1.0(2) MHz. The result of $\Delta \approx \nu$ shows that the system is not in the motional narrowing limit. The nonzero \textit{x} indicates that the spin-spin dynamical autocorrelation function $q(t) = (\tau/t)^x\mathrm{exp}(-\nu t)$ is not a simple exponential function. The fluctuation rate is about 2 orders of magnitude smaller than that of the canonical spin glass system \cite{Uemura1985}, and is comparable with several QSL candidates such as YbMgGaO$_4$ \cite{Li2016} and NaYbS$_2$ \cite{Sarkar-muon}, indicating that the spins are fluctuating collectively.

\section{Discussion and conclusions}\label{sec:5}
As was observed for YbMgGaO$_4$, the antisite disorder may be detrimental to the formation of the QSL state. From our structure refinement, the Al and Mg are randomly distributed at the 4\textit{f} site. However, careful considerations lead us to believe that it may not be as serious as it appears at first glance: The 4\textit{f} site is far from the magnetic layer, and there is no direct connection between the Al/MgO$_4$ tetrahedra and CeO$_{12}$ polyhedra. Therefore, the influence of the 4\textit{f} site mixing on the magnetic layer is likely negligible. In fact, the degree of disorder can be directly reflected in the magnetic susceptibility measurements, as evidenced by the spin-glass-like behavior observed in YbMgGaO$_4$ \cite{Ma2018}. In our sample, however, no sign of spin glass behavior is observed from ac susceptibility measurements.

The displacement from the 2\textit{d} to the 6\textit{h} site may be more serious since this is directly related to the magnetic ion positions. However, we note that the distance between the 2\textit{d} and 6\textit{h} sites is quite small, about 0.31 $\mathrm{\AA}$, compared to the ionic radius of Ce$^{3+}$ with 12 coordinates (1.34 $\mathrm{\AA}$) \cite{Shannon1976}. The Ce$^{3+}$ ions at the 6\textit{h} site should be also highly correlated, as the orphan spins are estimated to be only 0.2\% from the low temperature susceptibility tail. A further way to clarify the disorder effect is by inelastic neutron scattering measurement in the spin-polarized state. A recent study on CeMgAl$_{11}$O$_{19}$ shows that the magnon dispersion in the polarized state is rather sharp, reaching the instrumental resolution limit \cite{Gao2024}. All these results indicate that although there exist some degrees of disorder in CeMgAl$_{11}$O$_{19}$, their influence on the spin dynamics is likely negligible. Note that disorder is inevitable in real materials. Even for the so-called perfect triangular NaYbSe$_2$ system, careful structure refinement reveals about 5\% antisite disorder between Na and Yb \cite{Dai2021}. Thus, the ~13\% Ce site disorder is not an out-of-the-ordinarily high value for a real sample.

To conclude, we have identified the onset of antiferromagnetic correlations below $\sim$0.8 K for geometrically frustrated CeMgAl$_{11}$O$_{19}$. Moreover, neither long-range magnetic ordering nor a spin-glass-like transition has been observed down to 50 mK, indicating that the strong frustration inhibits the system's natural tendency to freeze, even at the lowest temperatures. The slow fluctuation frequency, 1.0(2) MHz at 50 mK, and the constant temperature dependence of the susceptibility all indicate that the spin-subsystem behaves collectively. The variation of magnetic specific heat with temperature and magnetic field supports the classification of
the title compound as a U(1) Dirac QSL state with dominant Ising anisotropy. Further neutron scattering measurements are attractive to clarify the spin excitations of the ground state.

\begin{acknowledgments}

We thank helpful discussions with Gang Chen and Shoushu Gong. This work is supported by the Guangdong Basic and Applied Basic Research Foundation (Grant No. 2022B1515120020) and National Natural Science Foundation of China with Grant No. 11874158. The $\mu$SR experiments were supported by MLF of J-PARC under a user program (proposal No. 2023B0022). Experiments at the ISIS Neutron and Muon Source were supported by a beamtime allocation RB2220102 from the Science and Technology Facilities Council. Data is available here: https://doi.org/10.5286/ISIS.E.RB2220102.

\end{acknowledgments}


\bibliography{ref}

\begin{thebibliography}{55}%
\makeatletter
\providecommand \@ifxundefined [1]{%
 \@ifx{#1\undefined}
}%
\providecommand \@ifnum [1]{%
 \ifnum #1\expandafter \@firstoftwo
 \else \expandafter \@secondoftwo
 \fi
}%
\providecommand \@ifx [1]{%
 \ifx #1\expandafter \@firstoftwo
 \else \expandafter \@secondoftwo
 \fi
}%
\providecommand \natexlab [1]{#1}%
\providecommand \enquote  [1]{``#1''}%
\providecommand \bibnamefont  [1]{#1}%
\providecommand \bibfnamefont [1]{#1}%
\providecommand \citenamefont [1]{#1}%
\providecommand \href@noop [0]{\@secondoftwo}%
\providecommand \href [0]{\begingroup \@sanitize@url \@href}%
\providecommand \@href[1]{\@@startlink{#1}\@@href}%
\providecommand \@@href[1]{\endgroup#1\@@endlink}%
\providecommand \@sanitize@url [0]{\catcode `\\12\catcode `\$12\catcode
  `\&12\catcode `\#12\catcode `\^12\catcode `\_12\catcode `\%12\relax}%
\providecommand \@@startlink[1]{}%
\providecommand \@@endlink[0]{}%
\providecommand \url  [0]{\begingroup\@sanitize@url \@url }%
\providecommand \@url [1]{\endgroup\@href {#1}{\urlprefix }}%
\providecommand \urlprefix  [0]{URL }%
\providecommand \Eprint [0]{\href }%
\providecommand \doibase [0]{http://dx.doi.org/}%
\providecommand \selectlanguage [0]{\@gobble}%
\providecommand \bibinfo  [0]{\@secondoftwo}%
\providecommand \bibfield  [0]{\@secondoftwo}%
\providecommand \translation [1]{[#1]}%
\providecommand \BibitemOpen [0]{}%
\providecommand \bibitemStop [0]{}%
\providecommand \bibitemNoStop [0]{.\EOS\space}%
\providecommand \EOS [0]{\spacefactor3000\relax}%
\providecommand \BibitemShut  [1]{\csname bibitem#1\endcsname}%
\let\auto@bib@innerbib\@empty
\bibitem [{\citenamefont {Heidarian}\ and\ \citenamefont
  {Paramekanti}(2010)}]{Heidarian2010}%
  \BibitemOpen
  \bibfield  {author} {\bibinfo {author} {\bibfnamefont {D.}~\bibnamefont
  {Heidarian}}\ and\ \bibinfo {author} {\bibfnamefont {A.}~\bibnamefont
  {Paramekanti}},\ }\href {\doibase 10.1103/PhysRevLett.104.015301} {\bibfield
  {journal} {\bibinfo  {journal} {Phys. Rev. Lett.}\ }\textbf {\bibinfo
  {volume} {104}},\ \bibinfo {pages} {015301} (\bibinfo {year}
  {2010})}\BibitemShut {NoStop}%
\bibitem [{\citenamefont {Gao}\ \emph {et~al.}(2022)\citenamefont {Gao},
  \citenamefont {Fan}, \citenamefont {Li}, \citenamefont {Yang}, \citenamefont
  {Zeng}, \citenamefont {Sheng}, \citenamefont {Zhong}, \citenamefont {Qi},
  \citenamefont {Wan},\ and\ \citenamefont {Li}}]{Gao2022}%
  \BibitemOpen
  \bibfield  {author} {\bibinfo {author} {\bibfnamefont {Y.}~\bibnamefont
  {Gao}}, \bibinfo {author} {\bibfnamefont {Y.-C.}\ \bibnamefont {Fan}},
  \bibinfo {author} {\bibfnamefont {H.}~\bibnamefont {Li}}, \bibinfo {author}
  {\bibfnamefont {F.}~\bibnamefont {Yang}}, \bibinfo {author} {\bibfnamefont
  {X.-T.}\ \bibnamefont {Zeng}}, \bibinfo {author} {\bibfnamefont {X.-L.}\
  \bibnamefont {Sheng}}, \bibinfo {author} {\bibfnamefont {R.}~\bibnamefont
  {Zhong}}, \bibinfo {author} {\bibfnamefont {Y.}~\bibnamefont {Qi}}, \bibinfo
  {author} {\bibfnamefont {Y.}~\bibnamefont {Wan}}, \ and\ \bibinfo {author}
  {\bibfnamefont {W.}~\bibnamefont {Li}},\ }\href {\doibase
  10.1038/s41535-022-00500-3} {\bibfield  {journal} {\bibinfo  {journal} {npj
  Quan. Mater.}\ }\textbf {\bibinfo {volume} {7}},\ \bibinfo {pages} {89}
  (\bibinfo {year} {2022})}\BibitemShut {NoStop}%
\bibitem [{\citenamefont {Xiang}\ \emph {et~al.}(2024)\citenamefont {Xiang},
  \citenamefont {Zhang}, \citenamefont {Gao}, \citenamefont {Schmidt},
  \citenamefont {Schmalzl}, \citenamefont {Wang}, \citenamefont {Li},
  \citenamefont {Xi}, \citenamefont {Liu}, \citenamefont {Jin}, \citenamefont
  {Li}, \citenamefont {Shen}, \citenamefont {Chen}, \citenamefont {Qi},
  \citenamefont {Wan}, \citenamefont {Jin}, \citenamefont {Li}, \citenamefont
  {Sun},\ and\ \citenamefont {Su}}]{Xiang2024}%
  \BibitemOpen
  \bibfield  {author} {\bibinfo {author} {\bibfnamefont {J.}~\bibnamefont
  {Xiang}}, \bibinfo {author} {\bibfnamefont {C.}~\bibnamefont {Zhang}},
  \bibinfo {author} {\bibfnamefont {Y.}~\bibnamefont {Gao}}, \bibinfo {author}
  {\bibfnamefont {W.}~\bibnamefont {Schmidt}}, \bibinfo {author} {\bibfnamefont
  {K.}~\bibnamefont {Schmalzl}}, \bibinfo {author} {\bibfnamefont {C.-W.}\
  \bibnamefont {Wang}}, \bibinfo {author} {\bibfnamefont {B.}~\bibnamefont
  {Li}}, \bibinfo {author} {\bibfnamefont {N.}~\bibnamefont {Xi}}, \bibinfo
  {author} {\bibfnamefont {X.-Y.}\ \bibnamefont {Liu}}, \bibinfo {author}
  {\bibfnamefont {H.}~\bibnamefont {Jin}}, \bibinfo {author} {\bibfnamefont
  {G.}~\bibnamefont {Li}}, \bibinfo {author} {\bibfnamefont {J.}~\bibnamefont
  {Shen}}, \bibinfo {author} {\bibfnamefont {Z.}~\bibnamefont {Chen}}, \bibinfo
  {author} {\bibfnamefont {Y.}~\bibnamefont {Qi}}, \bibinfo {author}
  {\bibfnamefont {Y.}~\bibnamefont {Wan}}, \bibinfo {author} {\bibfnamefont
  {W.}~\bibnamefont {Jin}}, \bibinfo {author} {\bibfnamefont {W.}~\bibnamefont
  {Li}}, \bibinfo {author} {\bibfnamefont {P.}~\bibnamefont {Sun}}, \ and\
  \bibinfo {author} {\bibfnamefont {G.}~\bibnamefont {Su}},\ }\href {\doibase
  10.1038/s41586-023-06885-w} {\bibfield  {journal} {\bibinfo  {journal}
  {Nature}\ }\textbf {\bibinfo {volume} {625}},\ \bibinfo {pages} {270}
  (\bibinfo {year} {2024})}\BibitemShut {NoStop}%
\bibitem [{\citenamefont {Balents}(2010)}]{Balents2010}%
  \BibitemOpen
  \bibfield  {author} {\bibinfo {author} {\bibfnamefont {L.}~\bibnamefont
  {Balents}},\ }\href {https://www.nature.com/articles/nature08917} {\bibfield
  {journal} {\bibinfo  {journal} {Nature}\ }\textbf {\bibinfo {volume} {464}},\
  \bibinfo {pages} {199} (\bibinfo {year} {2010})}\BibitemShut {NoStop}%
\bibitem [{\citenamefont {Savary}\ and\ \citenamefont
  {Balents}(2016)}]{Savary2016}%
  \BibitemOpen
  \bibfield  {author} {\bibinfo {author} {\bibfnamefont {L.}~\bibnamefont
  {Savary}}\ and\ \bibinfo {author} {\bibfnamefont {L.}~\bibnamefont
  {Balents}},\ }\href
  {https://iopscience.iop.org/article/10.1088/0034-4885/80/1/016502/meta}
  {\bibfield  {journal} {\bibinfo  {journal} {Rep. Prog. Phys.}\ }\textbf
  {\bibinfo {volume} {80}},\ \bibinfo {pages} {016502} (\bibinfo {year}
  {2016})}\BibitemShut {NoStop}%
\bibitem [{\citenamefont {Broholm}\ \emph {et~al.}(2020)\citenamefont
  {Broholm}, \citenamefont {Cava}, \citenamefont {Kivelson}, \citenamefont
  {Nocera}, \citenamefont {Norman},\ and\ \citenamefont
  {Senthil}}]{Broholm2020}%
  \BibitemOpen
  \bibfield  {author} {\bibinfo {author} {\bibfnamefont {C.}~\bibnamefont
  {Broholm}}, \bibinfo {author} {\bibfnamefont {R.}~\bibnamefont {Cava}},
  \bibinfo {author} {\bibfnamefont {S.}~\bibnamefont {Kivelson}}, \bibinfo
  {author} {\bibfnamefont {D.}~\bibnamefont {Nocera}}, \bibinfo {author}
  {\bibfnamefont {M.}~\bibnamefont {Norman}}, \ and\ \bibinfo {author}
  {\bibfnamefont {T.}~\bibnamefont {Senthil}},\ }\href
  {https://www.science.org/doi/abs/10.1126/science.aay0668} {\bibfield
  {journal} {\bibinfo  {journal} {Science}\ }\textbf {\bibinfo {volume}
  {367}},\ \bibinfo {pages} {eaay0668} (\bibinfo {year} {2020})}\BibitemShut
  {NoStop}%
\bibitem [{\citenamefont {Anderson}(1973)}]{Anderson1973}%
  \BibitemOpen
  \bibfield  {author} {\bibinfo {author} {\bibfnamefont {P.~W.}\ \bibnamefont
  {Anderson}},\ }\href {\doibase https://doi.org/10.1016/0025-5408(73)90167-0}
  {\bibfield  {journal} {\bibinfo  {journal} {Mat. Res. Bull.}\ }\textbf
  {\bibinfo {volume} {8}},\ \bibinfo {pages} {153} (\bibinfo {year}
  {1973})}\BibitemShut {NoStop}%
\bibitem [{\citenamefont {Anderson}(1987)}]{Anderson1987}%
  \BibitemOpen
  \bibfield  {author} {\bibinfo {author} {\bibfnamefont {P.~W.}\ \bibnamefont
  {Anderson}},\ }\href {\doibase
  https://www.science.org/doi/abs/10.1126/science.235.4793.1196} {\bibfield
  {journal} {\bibinfo  {journal} {Science}\ }\textbf {\bibinfo {volume}
  {235}},\ \bibinfo {pages} {1196} (\bibinfo {year} {1987})}\BibitemShut
  {NoStop}%
\bibitem [{\citenamefont {Mekata}\ \emph {et~al.}(1993)\citenamefont {Mekata},
  \citenamefont {Yaguchi}, \citenamefont {Takagi}, \citenamefont {Sugino},
  \citenamefont {Mitsuda}, \citenamefont {Yoshizawa}, \citenamefont {Hosoito},\
  and\ \citenamefont {Shinjo}}]{Mekata1993}%
  \BibitemOpen
  \bibfield  {author} {\bibinfo {author} {\bibfnamefont {M.}~\bibnamefont
  {Mekata}}, \bibinfo {author} {\bibfnamefont {N.}~\bibnamefont {Yaguchi}},
  \bibinfo {author} {\bibfnamefont {T.}~\bibnamefont {Takagi}}, \bibinfo
  {author} {\bibfnamefont {T.}~\bibnamefont {Sugino}}, \bibinfo {author}
  {\bibfnamefont {S.}~\bibnamefont {Mitsuda}}, \bibinfo {author} {\bibfnamefont
  {H.}~\bibnamefont {Yoshizawa}}, \bibinfo {author} {\bibfnamefont
  {N.}~\bibnamefont {Hosoito}}, \ and\ \bibinfo {author} {\bibfnamefont
  {T.}~\bibnamefont {Shinjo}},\ }\href {\doibase 10.1143/JPSJ.62.4474}
  {\bibfield  {journal} {\bibinfo  {journal} {J. Phys. Soc. Jpn.}\ }\textbf
  {\bibinfo {volume} {62}},\ \bibinfo {pages} {4474} (\bibinfo {year}
  {1993})}\BibitemShut {NoStop}%
\bibitem [{\citenamefont {Shirata}\ \emph {et~al.}(2012)\citenamefont
  {Shirata}, \citenamefont {Tanaka}, \citenamefont {Matsuo},\ and\
  \citenamefont {Kindo}}]{Shirata2012}%
  \BibitemOpen
  \bibfield  {author} {\bibinfo {author} {\bibfnamefont {Y.}~\bibnamefont
  {Shirata}}, \bibinfo {author} {\bibfnamefont {H.}~\bibnamefont {Tanaka}},
  \bibinfo {author} {\bibfnamefont {A.}~\bibnamefont {Matsuo}}, \ and\ \bibinfo
  {author} {\bibfnamefont {K.}~\bibnamefont {Kindo}},\ }\href {\doibase
  10.1103/PhysRevLett.108.057205} {\bibfield  {journal} {\bibinfo  {journal}
  {Phys. Rev. Lett.}\ }\textbf {\bibinfo {volume} {108}},\ \bibinfo {pages}
  {057205} (\bibinfo {year} {2012})}\BibitemShut {NoStop}%
\bibitem [{\citenamefont {Kimchi}\ \emph
  {et~al.}(2018{\natexlab{a}})\citenamefont {Kimchi}, \citenamefont {Nahum},\
  and\ \citenamefont {Senthil}}]{Kimchi2018}%
  \BibitemOpen
  \bibfield  {author} {\bibinfo {author} {\bibfnamefont {I.}~\bibnamefont
  {Kimchi}}, \bibinfo {author} {\bibfnamefont {A.}~\bibnamefont {Nahum}}, \
  and\ \bibinfo {author} {\bibfnamefont {T.}~\bibnamefont {Senthil}},\ }\href
  {\doibase 10.1103/PhysRevX.8.031028} {\bibfield  {journal} {\bibinfo
  {journal} {Phys. Rev. X}\ }\textbf {\bibinfo {volume} {8}},\ \bibinfo {pages}
  {031028} (\bibinfo {year} {2018}{\natexlab{a}})}\BibitemShut {NoStop}%
\bibitem [{\citenamefont {Kimchi}\ \emph
  {et~al.}(2018{\natexlab{b}})\citenamefont {Kimchi}, \citenamefont
  {Sheckelton}, \citenamefont {McQueen},\ and\ \citenamefont
  {Lee}}]{Kimchi2018-2}%
  \BibitemOpen
  \bibfield  {author} {\bibinfo {author} {\bibfnamefont {I.}~\bibnamefont
  {Kimchi}}, \bibinfo {author} {\bibfnamefont {J.~P.}\ \bibnamefont
  {Sheckelton}}, \bibinfo {author} {\bibfnamefont {T.~M.}\ \bibnamefont
  {McQueen}}, \ and\ \bibinfo {author} {\bibfnamefont {P.~A.}\ \bibnamefont
  {Lee}},\ }\href {\doibase 10.1038/s41467-018-06800-2} {\bibfield  {journal}
  {\bibinfo  {journal} {Nat. Commun.}\ }\textbf {\bibinfo {volume} {9}},\
  \bibinfo {pages} {4367} (\bibinfo {year} {2018}{\natexlab{b}})}\BibitemShut
  {NoStop}%
\bibitem [{\citenamefont {Ma}\ \emph {et~al.}(2018)\citenamefont {Ma},
  \citenamefont {Wang}, \citenamefont {Dong}, \citenamefont {Zhang},
  \citenamefont {Li}, \citenamefont {Zheng}, \citenamefont {Yu}, \citenamefont
  {Wang}, \citenamefont {Che}, \citenamefont {Ran}, \citenamefont {Bao},
  \citenamefont {Cai}, \citenamefont {\v{C}erm\'{a}k}, \citenamefont
  {Schneidewind}, \citenamefont {Yano}, \citenamefont {Gardner}, \citenamefont
  {Lu}, \citenamefont {Yu}, \citenamefont {Liu}, \citenamefont {Li},
  \citenamefont {Li},\ and\ \citenamefont {Wen}}]{Ma2018}%
  \BibitemOpen
  \bibfield  {author} {\bibinfo {author} {\bibfnamefont {Z.}~\bibnamefont
  {Ma}}, \bibinfo {author} {\bibfnamefont {J.}~\bibnamefont {Wang}}, \bibinfo
  {author} {\bibfnamefont {Z.-Y.}\ \bibnamefont {Dong}}, \bibinfo {author}
  {\bibfnamefont {J.}~\bibnamefont {Zhang}}, \bibinfo {author} {\bibfnamefont
  {S.}~\bibnamefont {Li}}, \bibinfo {author} {\bibfnamefont {S.-H.}\
  \bibnamefont {Zheng}}, \bibinfo {author} {\bibfnamefont {Y.}~\bibnamefont
  {Yu}}, \bibinfo {author} {\bibfnamefont {W.}~\bibnamefont {Wang}}, \bibinfo
  {author} {\bibfnamefont {L.}~\bibnamefont {Che}}, \bibinfo {author}
  {\bibfnamefont {K.}~\bibnamefont {Ran}}, \bibinfo {author} {\bibfnamefont
  {S.}~\bibnamefont {Bao}}, \bibinfo {author} {\bibfnamefont {Z.}~\bibnamefont
  {Cai}}, \bibinfo {author} {\bibfnamefont {P.}~\bibnamefont {\v{C}erm\'{a}k}},
  \bibinfo {author} {\bibfnamefont {A.}~\bibnamefont {Schneidewind}}, \bibinfo
  {author} {\bibfnamefont {S.}~\bibnamefont {Yano}}, \bibinfo {author}
  {\bibfnamefont {J.~S.}\ \bibnamefont {Gardner}}, \bibinfo {author}
  {\bibfnamefont {X.}~\bibnamefont {Lu}}, \bibinfo {author} {\bibfnamefont
  {S.-L.}\ \bibnamefont {Yu}}, \bibinfo {author} {\bibfnamefont {J.-M.}\
  \bibnamefont {Liu}}, \bibinfo {author} {\bibfnamefont {S.}~\bibnamefont
  {Li}}, \bibinfo {author} {\bibfnamefont {J.-X.}\ \bibnamefont {Li}}, \ and\
  \bibinfo {author} {\bibfnamefont {J.}~\bibnamefont {Wen}},\ }\href {\doibase
  10.1103/PhysRevLett.120.087201} {\bibfield  {journal} {\bibinfo  {journal}
  {Phys. Rev. Lett.}\ }\textbf {\bibinfo {volume} {120}},\ \bibinfo {pages}
  {087201} (\bibinfo {year} {2018})}\BibitemShut {NoStop}%
\bibitem [{\citenamefont {Li}(2019)}]{Li2019}%
  \BibitemOpen
  \bibfield  {author} {\bibinfo {author} {\bibfnamefont {Y.}~\bibnamefont
  {Li}},\ }\href {\doibase https://doi.org/10.1002/qute.201900089} {\bibfield
  {journal} {\bibinfo  {journal} {Adv. Quantum Techno.}\ }\textbf {\bibinfo
  {volume} {2}},\ \bibinfo {pages} {1900089} (\bibinfo {year}
  {2019})}\BibitemShut {NoStop}%
\bibitem [{\citenamefont {Li}\ \emph {et~al.}(2019)\citenamefont {Li},
  \citenamefont {Bachus}, \citenamefont {Liu}, \citenamefont {Radelytskyi},
  \citenamefont {Bertin}, \citenamefont {Schneidewind}, \citenamefont {Tokiwa},
  \citenamefont {Tsirlin},\ and\ \citenamefont {Gegenwart}}]{Li2019-2}%
  \BibitemOpen
  \bibfield  {author} {\bibinfo {author} {\bibfnamefont {Y.}~\bibnamefont
  {Li}}, \bibinfo {author} {\bibfnamefont {S.}~\bibnamefont {Bachus}}, \bibinfo
  {author} {\bibfnamefont {B.}~\bibnamefont {Liu}}, \bibinfo {author}
  {\bibfnamefont {I.}~\bibnamefont {Radelytskyi}}, \bibinfo {author}
  {\bibfnamefont {A.}~\bibnamefont {Bertin}}, \bibinfo {author} {\bibfnamefont
  {A.}~\bibnamefont {Schneidewind}}, \bibinfo {author} {\bibfnamefont
  {Y.}~\bibnamefont {Tokiwa}}, \bibinfo {author} {\bibfnamefont {A.~A.}\
  \bibnamefont {Tsirlin}}, \ and\ \bibinfo {author} {\bibfnamefont
  {P.}~\bibnamefont {Gegenwart}},\ }\href {\doibase
  10.1103/PhysRevLett.122.137201} {\bibfield  {journal} {\bibinfo  {journal}
  {Phys. Rev. Lett.}\ }\textbf {\bibinfo {volume} {122}},\ \bibinfo {pages}
  {137201} (\bibinfo {year} {2019})}\BibitemShut {NoStop}%
\bibitem [{\citenamefont {Wen}(2002)}]{Wen2002}%
  \BibitemOpen
  \bibfield  {author} {\bibinfo {author} {\bibfnamefont {X.-G.}\ \bibnamefont
  {Wen}},\ }\href {\doibase 10.1103/PhysRevB.65.165113} {\bibfield  {journal}
  {\bibinfo  {journal} {Phys. Rev. B}\ }\textbf {\bibinfo {volume} {65}},\
  \bibinfo {pages} {165113} (\bibinfo {year} {2002})}\BibitemShut {NoStop}%
\bibitem [{\citenamefont {Nagaosa}\ and\ \citenamefont
  {Lee}(1990)}]{Nagaosa1990}%
  \BibitemOpen
  \bibfield  {author} {\bibinfo {author} {\bibfnamefont {N.}~\bibnamefont
  {Nagaosa}}\ and\ \bibinfo {author} {\bibfnamefont {P.~A.}\ \bibnamefont
  {Lee}},\ }\href {\doibase 10.1103/PhysRevLett.64.2450} {\bibfield  {journal}
  {\bibinfo  {journal} {Phys. Rev. Lett.}\ }\textbf {\bibinfo {volume} {64}},\
  \bibinfo {pages} {2450} (\bibinfo {year} {1990})}\BibitemShut {NoStop}%
\bibitem [{\citenamefont {Ran}\ \emph {et~al.}(2007)\citenamefont {Ran},
  \citenamefont {Hermele}, \citenamefont {Lee},\ and\ \citenamefont
  {Wen}}]{Ran2007}%
  \BibitemOpen
  \bibfield  {author} {\bibinfo {author} {\bibfnamefont {Y.}~\bibnamefont
  {Ran}}, \bibinfo {author} {\bibfnamefont {M.}~\bibnamefont {Hermele}},
  \bibinfo {author} {\bibfnamefont {P.~A.}\ \bibnamefont {Lee}}, \ and\
  \bibinfo {author} {\bibfnamefont {X.-G.}\ \bibnamefont {Wen}},\ }\href
  {\doibase 10.1103/PhysRevLett.98.117205} {\bibfield  {journal} {\bibinfo
  {journal} {Phys. Rev. Lett.}\ }\textbf {\bibinfo {volume} {98}},\ \bibinfo
  {pages} {117205} (\bibinfo {year} {2007})}\BibitemShut {NoStop}%
\bibitem [{\citenamefont {Dai}\ \emph {et~al.}(2021)\citenamefont {Dai},
  \citenamefont {Zhang}, \citenamefont {Xie}, \citenamefont {Duan},
  \citenamefont {Gao}, \citenamefont {Zhu}, \citenamefont {Feng}, \citenamefont
  {Tao}, \citenamefont {Huang}, \citenamefont {Cao}, \citenamefont
  {Podlesnyak}, \citenamefont {Granroth}, \citenamefont {Everett},
  \citenamefont {Neuefeind}, \citenamefont {Voneshen}, \citenamefont {Wang},
  \citenamefont {Tan}, \citenamefont {Morosan}, \citenamefont {Wang},
  \citenamefont {Lin}, \citenamefont {Shu}, \citenamefont {Chen}, \citenamefont
  {Guo}, \citenamefont {Lu},\ and\ \citenamefont {Dai}}]{Dai2021}%
  \BibitemOpen
  \bibfield  {author} {\bibinfo {author} {\bibfnamefont {P.-L.}\ \bibnamefont
  {Dai}}, \bibinfo {author} {\bibfnamefont {G.}~\bibnamefont {Zhang}}, \bibinfo
  {author} {\bibfnamefont {Y.}~\bibnamefont {Xie}}, \bibinfo {author}
  {\bibfnamefont {C.}~\bibnamefont {Duan}}, \bibinfo {author} {\bibfnamefont
  {Y.}~\bibnamefont {Gao}}, \bibinfo {author} {\bibfnamefont {Z.}~\bibnamefont
  {Zhu}}, \bibinfo {author} {\bibfnamefont {E.}~\bibnamefont {Feng}}, \bibinfo
  {author} {\bibfnamefont {Z.}~\bibnamefont {Tao}}, \bibinfo {author}
  {\bibfnamefont {C.-L.}\ \bibnamefont {Huang}}, \bibinfo {author}
  {\bibfnamefont {H.}~\bibnamefont {Cao}}, \bibinfo {author} {\bibfnamefont
  {A.}~\bibnamefont {Podlesnyak}}, \bibinfo {author} {\bibfnamefont {G.~E.}\
  \bibnamefont {Granroth}}, \bibinfo {author} {\bibfnamefont {M.~S.}\
  \bibnamefont {Everett}}, \bibinfo {author} {\bibfnamefont {J.~C.}\
  \bibnamefont {Neuefeind}}, \bibinfo {author} {\bibfnamefont {D.}~\bibnamefont
  {Voneshen}}, \bibinfo {author} {\bibfnamefont {S.}~\bibnamefont {Wang}},
  \bibinfo {author} {\bibfnamefont {G.}~\bibnamefont {Tan}}, \bibinfo {author}
  {\bibfnamefont {E.}~\bibnamefont {Morosan}}, \bibinfo {author} {\bibfnamefont
  {X.}~\bibnamefont {Wang}}, \bibinfo {author} {\bibfnamefont {H.-Q.}\
  \bibnamefont {Lin}}, \bibinfo {author} {\bibfnamefont {L.}~\bibnamefont
  {Shu}}, \bibinfo {author} {\bibfnamefont {G.}~\bibnamefont {Chen}}, \bibinfo
  {author} {\bibfnamefont {Y.}~\bibnamefont {Guo}}, \bibinfo {author}
  {\bibfnamefont {X.}~\bibnamefont {Lu}}, \ and\ \bibinfo {author}
  {\bibfnamefont {P.}~\bibnamefont {Dai}},\ }\href {\doibase
  10.1103/PhysRevX.11.021044} {\bibfield  {journal} {\bibinfo  {journal} {Phys.
  Rev. X}\ }\textbf {\bibinfo {volume} {11}},\ \bibinfo {pages} {021044}
  (\bibinfo {year} {2021})}\BibitemShut {NoStop}%
\bibitem [{\citenamefont {Bag}\ \emph {et~al.}(2024)\citenamefont {Bag},
  \citenamefont {Xu}, \citenamefont {Sherman}, \citenamefont {Yadav},
  \citenamefont {Kolesnikov}, \citenamefont {Podlesnyak}, \citenamefont {Choi},
  \citenamefont {da~Silva}, \citenamefont {Moore},\ and\ \citenamefont
  {Haravifard}}]{Bag2024}%
  \BibitemOpen
  \bibfield  {author} {\bibinfo {author} {\bibfnamefont {R.}~\bibnamefont
  {Bag}}, \bibinfo {author} {\bibfnamefont {S.}~\bibnamefont {Xu}}, \bibinfo
  {author} {\bibfnamefont {N.~E.}\ \bibnamefont {Sherman}}, \bibinfo {author}
  {\bibfnamefont {L.}~\bibnamefont {Yadav}}, \bibinfo {author} {\bibfnamefont
  {A.~I.}\ \bibnamefont {Kolesnikov}}, \bibinfo {author} {\bibfnamefont
  {A.~A.}\ \bibnamefont {Podlesnyak}}, \bibinfo {author} {\bibfnamefont
  {E.~S.}\ \bibnamefont {Choi}}, \bibinfo {author} {\bibfnamefont
  {I.}~\bibnamefont {da~Silva}}, \bibinfo {author} {\bibfnamefont {J.~E.}\
  \bibnamefont {Moore}}, \ and\ \bibinfo {author} {\bibfnamefont
  {S.}~\bibnamefont {Haravifard}},\ }\href {\doibase
  10.1103/PhysRevLett.133.266703} {\bibfield  {journal} {\bibinfo  {journal}
  {Phys. Rev. Lett.}\ }\textbf {\bibinfo {volume} {133}},\ \bibinfo {pages}
  {266703} (\bibinfo {year} {2024})}\BibitemShut {NoStop}%
\bibitem [{\citenamefont {Ashtar}\ \emph {et~al.}(2019)\citenamefont {Ashtar},
  \citenamefont {Marwat}, \citenamefont {Gao}, \citenamefont {Zhang},
  \citenamefont {Pi}, \citenamefont {Yuan},\ and\ \citenamefont
  {Tian}}]{Ashtar2019}%
  \BibitemOpen
  \bibfield  {author} {\bibinfo {author} {\bibfnamefont {M.}~\bibnamefont
  {Ashtar}}, \bibinfo {author} {\bibfnamefont {M.~A.}\ \bibnamefont {Marwat}},
  \bibinfo {author} {\bibfnamefont {Y.~X.}\ \bibnamefont {Gao}}, \bibinfo
  {author} {\bibfnamefont {Z.~T.}\ \bibnamefont {Zhang}}, \bibinfo {author}
  {\bibfnamefont {L.}~\bibnamefont {Pi}}, \bibinfo {author} {\bibfnamefont
  {S.~L.}\ \bibnamefont {Yuan}}, \ and\ \bibinfo {author} {\bibfnamefont
  {Z.~M.}\ \bibnamefont {Tian}},\ }\href {\doibase 10.1039/C9TC02643F}
  {\bibfield  {journal} {\bibinfo  {journal} {J. Mater. Chem. C}\ }\textbf
  {\bibinfo {volume} {7}},\ \bibinfo {pages} {10073} (\bibinfo {year}
  {2019})}\BibitemShut {NoStop}%
\bibitem [{\citenamefont {Bu}\ \emph {et~al.}(2022)\citenamefont {Bu},
  \citenamefont {Ashtar}, \citenamefont {Shiroka}, \citenamefont {Walker},
  \citenamefont {Fu}, \citenamefont {Zhao}, \citenamefont {Gardner},
  \citenamefont {Chen}, \citenamefont {Tian},\ and\ \citenamefont
  {Guo}}]{Bu2022}%
  \BibitemOpen
  \bibfield  {author} {\bibinfo {author} {\bibfnamefont {H.}~\bibnamefont
  {Bu}}, \bibinfo {author} {\bibfnamefont {M.}~\bibnamefont {Ashtar}}, \bibinfo
  {author} {\bibfnamefont {T.}~\bibnamefont {Shiroka}}, \bibinfo {author}
  {\bibfnamefont {H.~C.}\ \bibnamefont {Walker}}, \bibinfo {author}
  {\bibfnamefont {Z.}~\bibnamefont {Fu}}, \bibinfo {author} {\bibfnamefont
  {J.}~\bibnamefont {Zhao}}, \bibinfo {author} {\bibfnamefont {J.~S.}\
  \bibnamefont {Gardner}}, \bibinfo {author} {\bibfnamefont {G.}~\bibnamefont
  {Chen}}, \bibinfo {author} {\bibfnamefont {Z.}~\bibnamefont {Tian}}, \ and\
  \bibinfo {author} {\bibfnamefont {H.}~\bibnamefont {Guo}},\ }\href {\doibase
  10.1103/PhysRevB.106.134428} {\bibfield  {journal} {\bibinfo  {journal}
  {Phys. Rev. B}\ }\textbf {\bibinfo {volume} {106}},\ \bibinfo {pages}
  {134428} (\bibinfo {year} {2022})}\BibitemShut {NoStop}%
\bibitem [{\citenamefont {Cao}\ \emph {et~al.}(2024)\citenamefont {Cao},
  \citenamefont {Bu}, \citenamefont {Fu}, \citenamefont {Zhao}, \citenamefont
  {Gardner}, \citenamefont {Ouyang}, \citenamefont {Tian}, \citenamefont {Li},\
  and\ \citenamefont {Guo}}]{Cao2024}%
  \BibitemOpen
  \bibfield  {author} {\bibinfo {author} {\bibfnamefont {Y.}~\bibnamefont
  {Cao}}, \bibinfo {author} {\bibfnamefont {H.}~\bibnamefont {Bu}}, \bibinfo
  {author} {\bibfnamefont {Z.}~\bibnamefont {Fu}}, \bibinfo {author}
  {\bibfnamefont {J.}~\bibnamefont {Zhao}}, \bibinfo {author} {\bibfnamefont
  {J.~S.}\ \bibnamefont {Gardner}}, \bibinfo {author} {\bibfnamefont
  {Z.}~\bibnamefont {Ouyang}}, \bibinfo {author} {\bibfnamefont
  {Z.}~\bibnamefont {Tian}}, \bibinfo {author} {\bibfnamefont {Z.}~\bibnamefont
  {Li}}, \ and\ \bibinfo {author} {\bibfnamefont {H.}~\bibnamefont {Guo}},\
  }\href {\doibase 10.1088/2752-5724/ad4a93} {\bibfield  {journal} {\bibinfo
  {journal} {Mater. Futures}\ }\textbf {\bibinfo {volume} {3}},\ \bibinfo
  {pages} {035201} (\bibinfo {year} {2024})}\BibitemShut {NoStop}%
\bibitem [{\citenamefont {Ma}\ \emph {et~al.}(2024)\citenamefont {Ma},
  \citenamefont {Zheng}, \citenamefont {Chen}, \citenamefont {Xu},
  \citenamefont {Dong}, \citenamefont {Wang}, \citenamefont {Du}, \citenamefont
  {Embs}, \citenamefont {Li}, \citenamefont {Li}, \citenamefont {Zhang},
  \citenamefont {Liu}, \citenamefont {Zhong}, \citenamefont {Liu},\ and\
  \citenamefont {Wen}}]{Ma2024}%
  \BibitemOpen
  \bibfield  {author} {\bibinfo {author} {\bibfnamefont {Z.}~\bibnamefont
  {Ma}}, \bibinfo {author} {\bibfnamefont {S.}~\bibnamefont {Zheng}}, \bibinfo
  {author} {\bibfnamefont {Y.}~\bibnamefont {Chen}}, \bibinfo {author}
  {\bibfnamefont {R.}~\bibnamefont {Xu}}, \bibinfo {author} {\bibfnamefont
  {Z.-Y.}\ \bibnamefont {Dong}}, \bibinfo {author} {\bibfnamefont
  {J.}~\bibnamefont {Wang}}, \bibinfo {author} {\bibfnamefont {H.}~\bibnamefont
  {Du}}, \bibinfo {author} {\bibfnamefont {J.~P.}\ \bibnamefont {Embs}},
  \bibinfo {author} {\bibfnamefont {S.}~\bibnamefont {Li}}, \bibinfo {author}
  {\bibfnamefont {Y.}~\bibnamefont {Li}}, \bibinfo {author} {\bibfnamefont
  {Y.}~\bibnamefont {Zhang}}, \bibinfo {author} {\bibfnamefont
  {M.}~\bibnamefont {Liu}}, \bibinfo {author} {\bibfnamefont {R.}~\bibnamefont
  {Zhong}}, \bibinfo {author} {\bibfnamefont {J.-M.}\ \bibnamefont {Liu}}, \
  and\ \bibinfo {author} {\bibfnamefont {J.}~\bibnamefont {Wen}},\ }\href
  {\doibase 10.1103/PhysRevB.109.165143} {\bibfield  {journal} {\bibinfo
  {journal} {Phys. Rev. B}\ }\textbf {\bibinfo {volume} {109}},\ \bibinfo
  {pages} {165143} (\bibinfo {year} {2024})}\BibitemShut {NoStop}%
\bibitem [{\citenamefont {Shen}\ \emph {et~al.}(2019)\citenamefont {Shen},
  \citenamefont {Liu}, \citenamefont {Qin}, \citenamefont {Shen}, \citenamefont
  {Li}, \citenamefont {Bewley}, \citenamefont {Schneidewind}, \citenamefont
  {Chen},\ and\ \citenamefont {Zhao}}]{Shen2019}%
  \BibitemOpen
  \bibfield  {author} {\bibinfo {author} {\bibfnamefont {Y.}~\bibnamefont
  {Shen}}, \bibinfo {author} {\bibfnamefont {C.}~\bibnamefont {Liu}}, \bibinfo
  {author} {\bibfnamefont {Y.}~\bibnamefont {Qin}}, \bibinfo {author}
  {\bibfnamefont {S.}~\bibnamefont {Shen}}, \bibinfo {author} {\bibfnamefont
  {Y.-D.}\ \bibnamefont {Li}}, \bibinfo {author} {\bibfnamefont
  {R.}~\bibnamefont {Bewley}}, \bibinfo {author} {\bibfnamefont
  {A.}~\bibnamefont {Schneidewind}}, \bibinfo {author} {\bibfnamefont
  {G.}~\bibnamefont {Chen}}, \ and\ \bibinfo {author} {\bibfnamefont
  {J.}~\bibnamefont {Zhao}},\ }\href {\doibase 10.1038/s41467-019-12410-3}
  {\bibfield  {journal} {\bibinfo  {journal} {Nat. Commun.}\ }\textbf {\bibinfo
  {volume} {10}},\ \bibinfo {pages} {4530} (\bibinfo {year}
  {2019})}\BibitemShut {NoStop}%
\bibitem [{\citenamefont {Chen}(2019)}]{chen2019}%
  \BibitemOpen
  \bibfield  {author} {\bibinfo {author} {\bibfnamefont {G.}~\bibnamefont
  {Chen}},\ }\href {\doibase 10.1103/PhysRevResearch.1.033141} {\bibfield
  {journal} {\bibinfo  {journal} {Phys. Rev. Res.}\ }\textbf {\bibinfo {volume}
  {1}},\ \bibinfo {pages} {033141} (\bibinfo {year} {2019})}\BibitemShut
  {NoStop}%
\bibitem [{\citenamefont {Pet\v{r}\'{\i}\v{c}ek}\ \emph
  {et~al.}(2014)\citenamefont {Pet\v{r}\'{\i}\v{c}ek}, \citenamefont
  {Du\v{s}ek},\ and\ \citenamefont {Palatinus}}]{Jana}%
  \BibitemOpen
  \bibfield  {author} {\bibinfo {author} {\bibfnamefont {V.}~\bibnamefont
  {Pet\v{r}\'{\i}\v{c}ek}}, \bibinfo {author} {\bibfnamefont {M.}~\bibnamefont
  {Du\v{s}ek}}, \ and\ \bibinfo {author} {\bibfnamefont {L.}~\bibnamefont
  {Palatinus}},\ }\href {\doibase doi:10.1515/zkri-2014-1737} {\bibfield
  {journal} {\bibinfo  {journal} {Z. Kristallogr.}\ }\textbf {\bibinfo {volume}
  {229}},\ \bibinfo {pages} {345} (\bibinfo {year} {2014})}\BibitemShut
  {NoStop}%
\bibitem [{\citenamefont {Rodr\'{\i}guez-Carvajal}(1993)}]{Fullprof}%
  \BibitemOpen
  \bibfield  {author} {\bibinfo {author} {\bibfnamefont {J.}~\bibnamefont
  {Rodr\'{\i}guez-Carvajal}},\ }\href {\doibase
  http://dx.doi.org/10.1016/0921-4526(93)90108-I} {\bibfield  {journal}
  {\bibinfo  {journal} {Physica B}\ }\textbf {\bibinfo {volume} {192}},\
  \bibinfo {pages} {55} (\bibinfo {year} {1993})}\BibitemShut {NoStop}%
\bibitem [{\citenamefont {Stevens}(1952)}]{Stevens1952}%
  \BibitemOpen
  \bibfield  {author} {\bibinfo {author} {\bibfnamefont {K.~W.~H.}\
  \bibnamefont {Stevens}},\ }\href {\doibase 10.1088/0370-1298/65/3/308}
  {\bibfield  {journal} {\bibinfo  {journal} {Proc. Phys. Soc. A}\ }\textbf
  {\bibinfo {volume} {65}},\ \bibinfo {pages} {209} (\bibinfo {year}
  {1952})}\BibitemShut {NoStop}%
\bibitem [{\citenamefont {Hutchings}(1964)}]{Hutchings1964}%
  \BibitemOpen
  \bibfield  {author} {\bibinfo {author} {\bibfnamefont {M.~T.}\ \bibnamefont
  {Hutchings}},\ }\href {\doibase
  https://doi.org/10.1016/S0081-1947(08)60517-2} {\bibfield  {journal}
  {\bibinfo  {journal} {Solid State Phys.}\ }\textbf {\bibinfo {volume} {16}},\
  \bibinfo {pages} {227} (\bibinfo {year} {1964})}\BibitemShut {NoStop}%
\bibitem [{\citenamefont {Scheie}(2021)}]{Scheie}%
  \BibitemOpen
  \bibfield  {author} {\bibinfo {author} {\bibfnamefont {A.}~\bibnamefont
  {Scheie}},\ }\href {\doibase 10.1107/S160057672001554X} {\bibfield  {journal}
  {\bibinfo  {journal} {J. Appl. Cryst.}\ }\textbf {\bibinfo {volume} {54}},\
  \bibinfo {pages} {356} (\bibinfo {year} {2021})}\BibitemShut {NoStop}%
\bibitem [{\citenamefont {Balz}\ \emph {et~al.}(2016)\citenamefont {Balz},
  \citenamefont {Lake}, \citenamefont {Reuther}, \citenamefont {Luetkens},
  \citenamefont {Sch\"{o}nemann}, \citenamefont {Herrmannsd\"{o}rfer},
  \citenamefont {Singh}, \citenamefont {Nazmul~Islam}, \citenamefont {Wheeler},
  \citenamefont {Rodriguez-Rivera}, \citenamefont {Guidi}, \citenamefont
  {Simeoni}, \citenamefont {Baines},\ and\ \citenamefont {Ryll}}]{Balz2016}%
  \BibitemOpen
  \bibfield  {author} {\bibinfo {author} {\bibfnamefont {C.}~\bibnamefont
  {Balz}}, \bibinfo {author} {\bibfnamefont {B.}~\bibnamefont {Lake}}, \bibinfo
  {author} {\bibfnamefont {J.}~\bibnamefont {Reuther}}, \bibinfo {author}
  {\bibfnamefont {H.}~\bibnamefont {Luetkens}}, \bibinfo {author}
  {\bibfnamefont {R.}~\bibnamefont {Sch\"{o}nemann}}, \bibinfo {author}
  {\bibfnamefont {T.}~\bibnamefont {Herrmannsd\"{o}rfer}}, \bibinfo {author}
  {\bibfnamefont {Y.}~\bibnamefont {Singh}}, \bibinfo {author} {\bibfnamefont
  {A.~T.~M.}\ \bibnamefont {Nazmul~Islam}}, \bibinfo {author} {\bibfnamefont
  {E.~M.}\ \bibnamefont {Wheeler}}, \bibinfo {author} {\bibfnamefont
  {J.}~\bibnamefont {Rodriguez-Rivera}}, \bibinfo {author} {\bibfnamefont
  {T.}~\bibnamefont {Guidi}}, \bibinfo {author} {\bibfnamefont
  {G.}~\bibnamefont {Simeoni}}, \bibinfo {author} {\bibfnamefont
  {C.}~\bibnamefont {Baines}}, \ and\ \bibinfo {author} {\bibfnamefont
  {H.}~\bibnamefont {Ryll}},\ }\href {\doibase 10.1038/nphys3826} {\bibfield
  {journal} {\bibinfo  {journal} {Nat. Phys.}\ }\textbf {\bibinfo {volume}
  {12}},\ \bibinfo {pages} {942} (\bibinfo {year} {2016})}\BibitemShut
  {NoStop}%
\bibitem [{\citenamefont {Gao}\ \emph {et~al.}(2024)\citenamefont {Gao},
  \citenamefont {Chen}, \citenamefont {Liu}, \citenamefont {Klemm},
  \citenamefont {Zhang}, \citenamefont {Ma}, \citenamefont {Xu}, \citenamefont
  {Won}, \citenamefont {McCandless}, \citenamefont {Murai}, \citenamefont
  {Ohira-Kawamura}, \citenamefont {Moxim}, \citenamefont {Ryan}, \citenamefont
  {Huang}, \citenamefont {Wang}, \citenamefont {Chan}, \citenamefont {Cheong},
  \citenamefont {Tchernyshyov}, \citenamefont {Balents},\ and\ \citenamefont
  {Dai}}]{Gao2024}%
  \BibitemOpen
  \bibfield  {author} {\bibinfo {author} {\bibfnamefont {B.}~\bibnamefont
  {Gao}}, \bibinfo {author} {\bibfnamefont {T.}~\bibnamefont {Chen}}, \bibinfo
  {author} {\bibfnamefont {C.}~\bibnamefont {Liu}}, \bibinfo {author}
  {\bibfnamefont {M.~L.}\ \bibnamefont {Klemm}}, \bibinfo {author}
  {\bibfnamefont {S.}~\bibnamefont {Zhang}}, \bibinfo {author} {\bibfnamefont
  {Z.}~\bibnamefont {Ma}}, \bibinfo {author} {\bibfnamefont {X.}~\bibnamefont
  {Xu}}, \bibinfo {author} {\bibfnamefont {C.}~\bibnamefont {Won}}, \bibinfo
  {author} {\bibfnamefont {T.}~\bibnamefont {McCandless}}, \bibinfo {author}
  {\bibfnamefont {N.}~\bibnamefont {Murai}}, \bibinfo {author} {\bibfnamefont
  {S.}~\bibnamefont {Ohira-Kawamura}}, \bibinfo {author} {\bibfnamefont
  {S.~J.}\ \bibnamefont {Moxim}}, \bibinfo {author} {\bibfnamefont {J.~T.}\
  \bibnamefont {Ryan}}, \bibinfo {author} {\bibfnamefont {X.}~\bibnamefont
  {Huang}}, \bibinfo {author} {\bibfnamefont {X.}~\bibnamefont {Wang}},
  \bibinfo {author} {\bibfnamefont {J.~Y.}\ \bibnamefont {Chan}}, \bibinfo
  {author} {\bibfnamefont {S.~W.}\ \bibnamefont {Cheong}}, \bibinfo {author}
  {\bibfnamefont {O.}~\bibnamefont {Tchernyshyov}}, \bibinfo {author}
  {\bibfnamefont {L.}~\bibnamefont {Balents}}, \ and\ \bibinfo {author}
  {\bibfnamefont {P.~C.}\ \bibnamefont {Dai}},\ }\href@noop {} {\bibfield
  {journal} {\bibinfo  {journal} {arXiv}\ }\textbf {\bibinfo {volume}
  {2408.15957}} (\bibinfo {year} {2024})}\BibitemShut {NoStop}%
\bibitem [{\citenamefont {Li}\ \emph {et~al.}(2016)\citenamefont {Li},
  \citenamefont {Adroja}, \citenamefont {Biswas}, \citenamefont {Baker},
  \citenamefont {Zhang}, \citenamefont {Liu}, \citenamefont {Tsirlin},
  \citenamefont {Gegenwart},\ and\ \citenamefont {Zhang}}]{Li2016}%
  \BibitemOpen
  \bibfield  {author} {\bibinfo {author} {\bibfnamefont {Y.}~\bibnamefont
  {Li}}, \bibinfo {author} {\bibfnamefont {D.}~\bibnamefont {Adroja}}, \bibinfo
  {author} {\bibfnamefont {P.~K.}\ \bibnamefont {Biswas}}, \bibinfo {author}
  {\bibfnamefont {P.~J.}\ \bibnamefont {Baker}}, \bibinfo {author}
  {\bibfnamefont {Q.}~\bibnamefont {Zhang}}, \bibinfo {author} {\bibfnamefont
  {J.}~\bibnamefont {Liu}}, \bibinfo {author} {\bibfnamefont {A.~A.}\
  \bibnamefont {Tsirlin}}, \bibinfo {author} {\bibfnamefont {P.}~\bibnamefont
  {Gegenwart}}, \ and\ \bibinfo {author} {\bibfnamefont {Q.}~\bibnamefont
  {Zhang}},\ }\href {\doibase 10.1103/PhysRevLett.117.097201} {\bibfield
  {journal} {\bibinfo  {journal} {Phys. Rev. Lett.}\ }\textbf {\bibinfo
  {volume} {117}},\ \bibinfo {pages} {097201} (\bibinfo {year}
  {2016})}\BibitemShut {NoStop}%
\bibitem [{\citenamefont {Isono}\ \emph {et~al.}(2014)\citenamefont {Isono},
  \citenamefont {Kamo}, \citenamefont {Ueda}, \citenamefont {Takahashi},
  \citenamefont {Kimata}, \citenamefont {Tajima}, \citenamefont {Tsuchiya},
  \citenamefont {Terashima}, \citenamefont {Uji},\ and\ \citenamefont
  {Mori}}]{Isono2014}%
  \BibitemOpen
  \bibfield  {author} {\bibinfo {author} {\bibfnamefont {T.}~\bibnamefont
  {Isono}}, \bibinfo {author} {\bibfnamefont {H.}~\bibnamefont {Kamo}},
  \bibinfo {author} {\bibfnamefont {A.}~\bibnamefont {Ueda}}, \bibinfo {author}
  {\bibfnamefont {K.}~\bibnamefont {Takahashi}}, \bibinfo {author}
  {\bibfnamefont {M.}~\bibnamefont {Kimata}}, \bibinfo {author} {\bibfnamefont
  {H.}~\bibnamefont {Tajima}}, \bibinfo {author} {\bibfnamefont
  {S.}~\bibnamefont {Tsuchiya}}, \bibinfo {author} {\bibfnamefont
  {T.}~\bibnamefont {Terashima}}, \bibinfo {author} {\bibfnamefont
  {S.}~\bibnamefont {Uji}}, \ and\ \bibinfo {author} {\bibfnamefont
  {H.}~\bibnamefont {Mori}},\ }\href {\doibase 10.1103/PhysRevLett.112.177201}
  {\bibfield  {journal} {\bibinfo  {journal} {Phys. Rev. Lett.}\ }\textbf
  {\bibinfo {volume} {112}},\ \bibinfo {pages} {177201} (\bibinfo {year}
  {2014})}\BibitemShut {NoStop}%
\bibitem [{\citenamefont {Gopal}(1966)}]{Gopal}%
  \BibitemOpen
  \bibfield  {author} {\bibinfo {author} {\bibfnamefont {E.}~\bibnamefont
  {Gopal}},\ }\href@noop {} {\emph {\bibinfo {title} {{Sepcific Heats At Low
  Temperatures}}}}\ (\bibinfo  {publisher} {Plenum Press},\ \bibinfo {address}
  {New York},\ \bibinfo {year} {1966})\BibitemShut {NoStop}%
\bibitem [{\citenamefont {Motrunich}(2005)}]{Motrunich2005}%
  \BibitemOpen
  \bibfield  {author} {\bibinfo {author} {\bibfnamefont {O.~I.}\ \bibnamefont
  {Motrunich}},\ }\href {\doibase 10.1103/PhysRevB.72.045105} {\bibfield
  {journal} {\bibinfo  {journal} {Phys. Rev. B}\ }\textbf {\bibinfo {volume}
  {72}},\ \bibinfo {pages} {045105} (\bibinfo {year} {2005})}\BibitemShut
  {NoStop}%
\bibitem [{\citenamefont {Yamashita}\ \emph {et~al.}(2008)\citenamefont
  {Yamashita}, \citenamefont {Nakazawa}, \citenamefont {Oguni}, \citenamefont
  {Oshima}, \citenamefont {Nojiri}, \citenamefont {Shimizu}, \citenamefont
  {Miyagawa},\ and\ \citenamefont {Kanoda}}]{Yamashita2008}%
  \BibitemOpen
  \bibfield  {author} {\bibinfo {author} {\bibfnamefont {S.}~\bibnamefont
  {Yamashita}}, \bibinfo {author} {\bibfnamefont {Y.}~\bibnamefont {Nakazawa}},
  \bibinfo {author} {\bibfnamefont {M.}~\bibnamefont {Oguni}}, \bibinfo
  {author} {\bibfnamefont {Y.}~\bibnamefont {Oshima}}, \bibinfo {author}
  {\bibfnamefont {H.}~\bibnamefont {Nojiri}}, \bibinfo {author} {\bibfnamefont
  {Y.}~\bibnamefont {Shimizu}}, \bibinfo {author} {\bibfnamefont
  {K.}~\bibnamefont {Miyagawa}}, \ and\ \bibinfo {author} {\bibfnamefont
  {K.}~\bibnamefont {Kanoda}},\ }\href {\doibase 10.1038/nphys942} {\bibfield
  {journal} {\bibinfo  {journal} {Nat. Phys.}\ }\textbf {\bibinfo {volume}
  {4}},\ \bibinfo {pages} {459} (\bibinfo {year} {2008})}\BibitemShut {NoStop}%
\bibitem [{\citenamefont {Savary}\ and\ \citenamefont
  {Balents}(2012)}]{Savary2012}%
  \BibitemOpen
  \bibfield  {author} {\bibinfo {author} {\bibfnamefont {L.}~\bibnamefont
  {Savary}}\ and\ \bibinfo {author} {\bibfnamefont {L.}~\bibnamefont
  {Balents}},\ }\href {\doibase 10.1103/PhysRevLett.108.037202} {\bibfield
  {journal} {\bibinfo  {journal} {Phys. Rev. Lett.}\ }\textbf {\bibinfo
  {volume} {108}},\ \bibinfo {pages} {037202} (\bibinfo {year}
  {2012})}\BibitemShut {NoStop}%
\bibitem [{\citenamefont {Guo}\ \emph {et~al.}(2013)\citenamefont {Guo},
  \citenamefont {Tanida}, \citenamefont {Kobayashi}, \citenamefont {Kawasaki},
  \citenamefont {Sera}, \citenamefont {Nishioka}, \citenamefont {Matsumura},
  \citenamefont {Watanabe},\ and\ \citenamefont {Xu}}]{Guo2013}%
  \BibitemOpen
  \bibfield  {author} {\bibinfo {author} {\bibfnamefont {H.}~\bibnamefont
  {Guo}}, \bibinfo {author} {\bibfnamefont {H.}~\bibnamefont {Tanida}},
  \bibinfo {author} {\bibfnamefont {R.}~\bibnamefont {Kobayashi}}, \bibinfo
  {author} {\bibfnamefont {I.}~\bibnamefont {Kawasaki}}, \bibinfo {author}
  {\bibfnamefont {M.}~\bibnamefont {Sera}}, \bibinfo {author} {\bibfnamefont
  {T.}~\bibnamefont {Nishioka}}, \bibinfo {author} {\bibfnamefont
  {M.}~\bibnamefont {Matsumura}}, \bibinfo {author} {\bibfnamefont
  {I.}~\bibnamefont {Watanabe}}, \ and\ \bibinfo {author} {\bibfnamefont
  {Z.-a.}\ \bibnamefont {Xu}},\ }\href {\doibase 10.1103/PhysRevB.88.115206}
  {\bibfield  {journal} {\bibinfo  {journal} {Phys. Rev. B}\ }\textbf {\bibinfo
  {volume} {88}},\ \bibinfo {pages} {115206} (\bibinfo {year}
  {2013})}\BibitemShut {NoStop}%
\bibitem [{\citenamefont {Ishant}\ \emph {et~al.}(2024)\citenamefont {Ishant},
  \citenamefont {Shiroka}, \citenamefont {Stockert}, \citenamefont {Fritsch},\
  and\ \citenamefont {Majumder}}]{Ishant2024}%
  \BibitemOpen
  \bibfield  {author} {\bibinfo {author} {\bibfnamefont {I.}~\bibnamefont
  {Ishant}}, \bibinfo {author} {\bibfnamefont {T.}~\bibnamefont {Shiroka}},
  \bibinfo {author} {\bibfnamefont {O.}~\bibnamefont {Stockert}}, \bibinfo
  {author} {\bibfnamefont {V.}~\bibnamefont {Fritsch}}, \ and\ \bibinfo
  {author} {\bibfnamefont {M.}~\bibnamefont {Majumder}},\ }\href {\doibase
  10.1103/PhysRevResearch.6.023112} {\bibfield  {journal} {\bibinfo  {journal}
  {Phys. Rev. Res.}\ }\textbf {\bibinfo {volume} {6}},\ \bibinfo {pages}
  {023112} (\bibinfo {year} {2024})}\BibitemShut {NoStop}%
\bibitem [{\citenamefont {Campbell}\ \emph {et~al.}(1994)\citenamefont
  {Campbell}, \citenamefont {Amato}, \citenamefont {Gygax}, \citenamefont
  {Herlach}, \citenamefont {Schenck}, \citenamefont {Cywinski},\ and\
  \citenamefont {Kilcoyne}}]{Campbell1994}%
  \BibitemOpen
  \bibfield  {author} {\bibinfo {author} {\bibfnamefont {I.~A.}\ \bibnamefont
  {Campbell}}, \bibinfo {author} {\bibfnamefont {A.}~\bibnamefont {Amato}},
  \bibinfo {author} {\bibfnamefont {F.~N.}\ \bibnamefont {Gygax}}, \bibinfo
  {author} {\bibfnamefont {D.}~\bibnamefont {Herlach}}, \bibinfo {author}
  {\bibfnamefont {A.}~\bibnamefont {Schenck}}, \bibinfo {author} {\bibfnamefont
  {R.}~\bibnamefont {Cywinski}}, \ and\ \bibinfo {author} {\bibfnamefont
  {S.~H.}\ \bibnamefont {Kilcoyne}},\ }\href
  {http://link.aps.org/doi/10.1103/PhysRevLett.72.1291} {\bibfield  {journal}
  {\bibinfo  {journal} {Phys. Rev. Lett.}\ }\textbf {\bibinfo {volume} {72}},\
  \bibinfo {pages} {1291} (\bibinfo {year} {1994})}\BibitemShut {NoStop}%
\bibitem [{\citenamefont {Keren}\ \emph {et~al.}(1996)\citenamefont {Keren},
  \citenamefont {Mendels}, \citenamefont {Campbell},\ and\ \citenamefont
  {Lord}}]{Keren1996}%
  \BibitemOpen
  \bibfield  {author} {\bibinfo {author} {\bibfnamefont {A.}~\bibnamefont
  {Keren}}, \bibinfo {author} {\bibfnamefont {P.}~\bibnamefont {Mendels}},
  \bibinfo {author} {\bibfnamefont {I.~A.}\ \bibnamefont {Campbell}}, \ and\
  \bibinfo {author} {\bibfnamefont {J.}~\bibnamefont {Lord}},\ }\href
  {http://link.aps.org/doi/10.1103/PhysRevLett.77.1386} {\bibfield  {journal}
  {\bibinfo  {journal} {Phys. Rev. Lett.}\ }\textbf {\bibinfo {volume} {77}},\
  \bibinfo {pages} {1386} (\bibinfo {year} {1996})}\BibitemShut {NoStop}%
\bibitem [{\citenamefont {Guo}\ \emph {et~al.}(2014)\citenamefont {Guo},
  \citenamefont {Xing}, \citenamefont {Tong}, \citenamefont {Tao},
  \citenamefont {Watanabe},\ and\ \citenamefont {Xu}}]{Guo2014}%
  \BibitemOpen
  \bibfield  {author} {\bibinfo {author} {\bibfnamefont {H.}~\bibnamefont
  {Guo}}, \bibinfo {author} {\bibfnamefont {H.}~\bibnamefont {Xing}}, \bibinfo
  {author} {\bibfnamefont {J.}~\bibnamefont {Tong}}, \bibinfo {author}
  {\bibfnamefont {Q.}~\bibnamefont {Tao}}, \bibinfo {author} {\bibfnamefont
  {I.}~\bibnamefont {Watanabe}}, \ and\ \bibinfo {author} {\bibfnamefont
  {Z.-a.}\ \bibnamefont {Xu}},\ }\href {\doibase
  10.1088/0953-8984/26/43/436002} {\bibfield  {journal} {\bibinfo  {journal}
  {J. Phys.: Condens. Matter}\ }\textbf {\bibinfo {volume} {26}},\ \bibinfo
  {pages} {436002} (\bibinfo {year} {2014})}\BibitemShut {NoStop}%
\bibitem [{\citenamefont {Keren}\ \emph {et~al.}(2001)\citenamefont {Keren},
  \citenamefont {Bazalitsky}, \citenamefont {Campbell},\ and\ \citenamefont
  {Lord}}]{Keren2001}%
  \BibitemOpen
  \bibfield  {author} {\bibinfo {author} {\bibfnamefont {A.}~\bibnamefont
  {Keren}}, \bibinfo {author} {\bibfnamefont {G.}~\bibnamefont {Bazalitsky}},
  \bibinfo {author} {\bibfnamefont {I.}~\bibnamefont {Campbell}}, \ and\
  \bibinfo {author} {\bibfnamefont {J.~S.}\ \bibnamefont {Lord}},\ }\href
  {\doibase 10.1103/PhysRevB.64.054403} {\bibfield  {journal} {\bibinfo
  {journal} {Phys. Rev. B}\ }\textbf {\bibinfo {volume} {64}},\ \bibinfo
  {pages} {054403} (\bibinfo {year} {2001})}\BibitemShut {NoStop}%
\bibitem [{\citenamefont {Uemura}\ \emph {et~al.}(1985)\citenamefont {Uemura},
  \citenamefont {Yamazaki}, \citenamefont {Harshman}, \citenamefont {Senba},\
  and\ \citenamefont {Ansaldo}}]{Uemura1985}%
  \BibitemOpen
  \bibfield  {author} {\bibinfo {author} {\bibfnamefont {Y.~J.}\ \bibnamefont
  {Uemura}}, \bibinfo {author} {\bibfnamefont {T.}~\bibnamefont {Yamazaki}},
  \bibinfo {author} {\bibfnamefont {D.~R.}\ \bibnamefont {Harshman}}, \bibinfo
  {author} {\bibfnamefont {M.}~\bibnamefont {Senba}}, \ and\ \bibinfo {author}
  {\bibfnamefont {E.~J.}\ \bibnamefont {Ansaldo}},\ }\href
  {http://link.aps.org/doi/10.1103/PhysRevB.31.546} {\bibfield  {journal}
  {\bibinfo  {journal} {Phys. Rev. B}\ }\textbf {\bibinfo {volume} {31}},\
  \bibinfo {pages} {546} (\bibinfo {year} {1985})}\BibitemShut {NoStop}%
\bibitem [{\citenamefont {Sarkar}\ \emph {et~al.}(2019)\citenamefont {Sarkar},
  \citenamefont {Schlender}, \citenamefont {Grinenko}, \citenamefont
  {Haeussler}, \citenamefont {Baker}, \citenamefont {Doert},\ and\
  \citenamefont {Klauss}}]{Sarkar-muon}%
  \BibitemOpen
  \bibfield  {author} {\bibinfo {author} {\bibfnamefont {R.}~\bibnamefont
  {Sarkar}}, \bibinfo {author} {\bibfnamefont {P.}~\bibnamefont {Schlender}},
  \bibinfo {author} {\bibfnamefont {V.}~\bibnamefont {Grinenko}}, \bibinfo
  {author} {\bibfnamefont {E.}~\bibnamefont {Haeussler}}, \bibinfo {author}
  {\bibfnamefont {P.~J.}\ \bibnamefont {Baker}}, \bibinfo {author}
  {\bibfnamefont {T.}~\bibnamefont {Doert}}, \ and\ \bibinfo {author}
  {\bibfnamefont {H.~H.}\ \bibnamefont {Klauss}},\ }\href {\doibase
  10.1103/PhysRevB.100.241116} {\bibfield  {journal} {\bibinfo  {journal}
  {Phys. Rev. B}\ }\textbf {\bibinfo {volume} {100}},\ \bibinfo {pages}
  {241116} (\bibinfo {year} {2019})}\BibitemShut {NoStop}%
\bibitem [{\citenamefont {Shannon}(1976)}]{Shannon1976}%
  \BibitemOpen
  \bibfield  {author} {\bibinfo {author} {\bibfnamefont {R.~D.}\ \bibnamefont
  {Shannon}},\ }\href {\doibase 10.1107/S0567739476001551} {\bibfield
  {journal} {\bibinfo  {journal} {Acta Cryst. A}\ }\textbf {\bibinfo {volume}
  {32}},\ \bibinfo {pages} {751} (\bibinfo {year} {1976})}\BibitemShut
  {NoStop}%
\bibitem [{\citenamefont {B{\^e}che}\ \emph {et~al.}(2008)\citenamefont
  {B{\^e}che}, \citenamefont {Charvin}, \citenamefont {Perarnau}, \citenamefont
  {Abanades},\ and\ \citenamefont {Flamant}}]{beche20083d}%
  \BibitemOpen
  \bibfield  {author} {\bibinfo {author} {\bibfnamefont {E.}~\bibnamefont
  {B{\^e}che}}, \bibinfo {author} {\bibfnamefont {P.}~\bibnamefont {Charvin}},
  \bibinfo {author} {\bibfnamefont {D.}~\bibnamefont {Perarnau}}, \bibinfo
  {author} {\bibfnamefont {S.}~\bibnamefont {Abanades}}, \ and\ \bibinfo
  {author} {\bibfnamefont {G.}~\bibnamefont {Flamant}},\ }\href
  {https://analyticalsciencejournals.onlinelibrary.wiley.com/doi/abs/10.1002/sia.2686}
  {\bibfield  {journal} {\bibinfo  {journal} {Surf. Interface Anal.}\ }\textbf
  {\bibinfo {volume} {40}},\ \bibinfo {pages} {264} (\bibinfo {year}
  {2008})}\BibitemShut {NoStop}%
\bibitem [{\citenamefont {Teterin}\ \emph {et~al.}(1998)\citenamefont
  {Teterin}, \citenamefont {Teterin}, \citenamefont {Lebedev},\ and\
  \citenamefont {Utkin}}]{teterin1998xps}%
  \BibitemOpen
  \bibfield  {author} {\bibinfo {author} {\bibfnamefont {Y.~A.}\ \bibnamefont
  {Teterin}}, \bibinfo {author} {\bibfnamefont {A.~Y.}\ \bibnamefont
  {Teterin}}, \bibinfo {author} {\bibfnamefont {A.}~\bibnamefont {Lebedev}}, \
  and\ \bibinfo {author} {\bibfnamefont {I.}~\bibnamefont {Utkin}},\ }\href
  {https://www.sciencedirect.com/science/article/pii/S0368204897001394}
  {\bibfield  {journal} {\bibinfo  {journal} {J. Electron Spectrosc.}\ }\textbf
  {\bibinfo {volume} {88}},\ \bibinfo {pages} {275} (\bibinfo {year}
  {1998})}\BibitemShut {NoStop}%
\bibitem [{\citenamefont {Gurgul}\ \emph {et~al.}(2013)\citenamefont {Gurgul},
  \citenamefont {Rinke}, \citenamefont {Schellenberg},\ and\ \citenamefont
  {P{\"o}ttgen}}]{gurgul2013antimonide}%
  \BibitemOpen
  \bibfield  {author} {\bibinfo {author} {\bibfnamefont {J.}~\bibnamefont
  {Gurgul}}, \bibinfo {author} {\bibfnamefont {M.~T.}\ \bibnamefont {Rinke}},
  \bibinfo {author} {\bibfnamefont {I.}~\bibnamefont {Schellenberg}}, \ and\
  \bibinfo {author} {\bibfnamefont {R.}~\bibnamefont {P{\"o}ttgen}},\ }\href
  {https://www.sciencedirect.com/science/article/pii/S1293255812003779}
  {\bibfield  {journal} {\bibinfo  {journal} {Solid State sci.}\ }\textbf
  {\bibinfo {volume} {17}},\ \bibinfo {pages} {122} (\bibinfo {year}
  {2013})}\BibitemShut {NoStop}%
\bibitem [{\citenamefont {Gaudet}\ \emph {et~al.}(2019)\citenamefont {Gaudet},
  \citenamefont {Smith}, \citenamefont {Dudemaine}, \citenamefont {Beare},
  \citenamefont {Buhariwalla}, \citenamefont {Butch}, \citenamefont {Stone},
  \citenamefont {Kolesnikov}, \citenamefont {Xu}, \citenamefont {Yahne},
  \citenamefont {Ross}, \citenamefont {Marjerrison}, \citenamefont {Garrett},
  \citenamefont {Luke}, \citenamefont {Bianchi},\ and\ \citenamefont
  {Gaulin}}]{Gaudet2019}%
  \BibitemOpen
  \bibfield  {author} {\bibinfo {author} {\bibfnamefont {J.}~\bibnamefont
  {Gaudet}}, \bibinfo {author} {\bibfnamefont {E.~M.}\ \bibnamefont {Smith}},
  \bibinfo {author} {\bibfnamefont {J.}~\bibnamefont {Dudemaine}}, \bibinfo
  {author} {\bibfnamefont {J.}~\bibnamefont {Beare}}, \bibinfo {author}
  {\bibfnamefont {C.~R.~C.}\ \bibnamefont {Buhariwalla}}, \bibinfo {author}
  {\bibfnamefont {N.~P.}\ \bibnamefont {Butch}}, \bibinfo {author}
  {\bibfnamefont {M.~B.}\ \bibnamefont {Stone}}, \bibinfo {author}
  {\bibfnamefont {A.~I.}\ \bibnamefont {Kolesnikov}}, \bibinfo {author}
  {\bibfnamefont {G.}~\bibnamefont {Xu}}, \bibinfo {author} {\bibfnamefont
  {D.~R.}\ \bibnamefont {Yahne}}, \bibinfo {author} {\bibfnamefont {K.~A.}\
  \bibnamefont {Ross}}, \bibinfo {author} {\bibfnamefont {C.~A.}\ \bibnamefont
  {Marjerrison}}, \bibinfo {author} {\bibfnamefont {J.~D.}\ \bibnamefont
  {Garrett}}, \bibinfo {author} {\bibfnamefont {G.~M.}\ \bibnamefont {Luke}},
  \bibinfo {author} {\bibfnamefont {A.~D.}\ \bibnamefont {Bianchi}}, \ and\
  \bibinfo {author} {\bibfnamefont {B.~D.}\ \bibnamefont {Gaulin}},\ }\href
  {\doibase 10.1103/PhysRevLett.122.187201} {\bibfield  {journal} {\bibinfo
  {journal} {Phys. Rev. Lett.}\ }\textbf {\bibinfo {volume} {122}},\ \bibinfo
  {pages} {187201} (\bibinfo {year} {2019})}\BibitemShut {NoStop}%
\bibitem [{\citenamefont {Kresse}\ and\ \citenamefont
  {Hafner}(1993)}]{Kresse1993}%
  \BibitemOpen
  \bibfield  {author} {\bibinfo {author} {\bibfnamefont {G.}~\bibnamefont
  {Kresse}}\ and\ \bibinfo {author} {\bibfnamefont {J.}~\bibnamefont
  {Hafner}},\ }\href {\doibase 10.1103/PhysRevB.47.558} {\bibfield  {journal}
  {\bibinfo  {journal} {Phys. Rev. B}\ }\textbf {\bibinfo {volume} {47}},\
  \bibinfo {pages} {558} (\bibinfo {year} {1993})}\BibitemShut {NoStop}%
\bibitem [{\citenamefont {Kresse}\ and\ \citenamefont
  {Furthm\"{u}ller}(1996)}]{Kresse1996}%
  \BibitemOpen
  \bibfield  {author} {\bibinfo {author} {\bibfnamefont {G.}~\bibnamefont
  {Kresse}}\ and\ \bibinfo {author} {\bibfnamefont {J.}~\bibnamefont
  {Furthm\"{u}ller}},\ }\href {\doibase
  https://doi.org/10.1016/0927-0256(96)00008-0} {\bibfield  {journal} {\bibinfo
   {journal} {Comput. Mater. Sci.}\ }\textbf {\bibinfo {volume} {6}},\ \bibinfo
  {pages} {15} (\bibinfo {year} {1996})}\BibitemShut {NoStop}%
\bibitem [{\citenamefont {Abragam}\ and\ \citenamefont
  {Bleaney}(1970)}]{Abragam}%
  \BibitemOpen
  \bibfield  {author} {\bibinfo {author} {\bibfnamefont {A.}~\bibnamefont
  {Abragam}}\ and\ \bibinfo {author} {\bibfnamefont {B.}~\bibnamefont
  {Bleaney}},\ }\href@noop {} {\emph {\bibinfo {title} {Electron Paramagnetic
  Resonance of Transition Ions}}},\ edited by\ \bibinfo {editor} {\bibfnamefont
  {W.}~\bibnamefont {Marshall}}\ and\ \bibinfo {editor} {\bibfnamefont {D.~H.}\
  \bibnamefont {Wilkinson}}\ (\bibinfo  {publisher} {Clarendon Press},\
  \bibinfo {address} {Oxford},\ \bibinfo {year} {1970})\BibitemShut {NoStop}%
\end{thebibliography}%

\clearpage
\setcounter{figure}{0}
\setcounter{table}{0}
\renewcommand{\thefigure}{S\arabic{figure}}
\renewcommand{\thetable}{S\arabic{table}}

\section*{Supplementary Materials}
\textbf{1. Crystal structure refinement}

\begin{figure}[h]
  \centering
  \includegraphics[width=0.9\columnwidth]{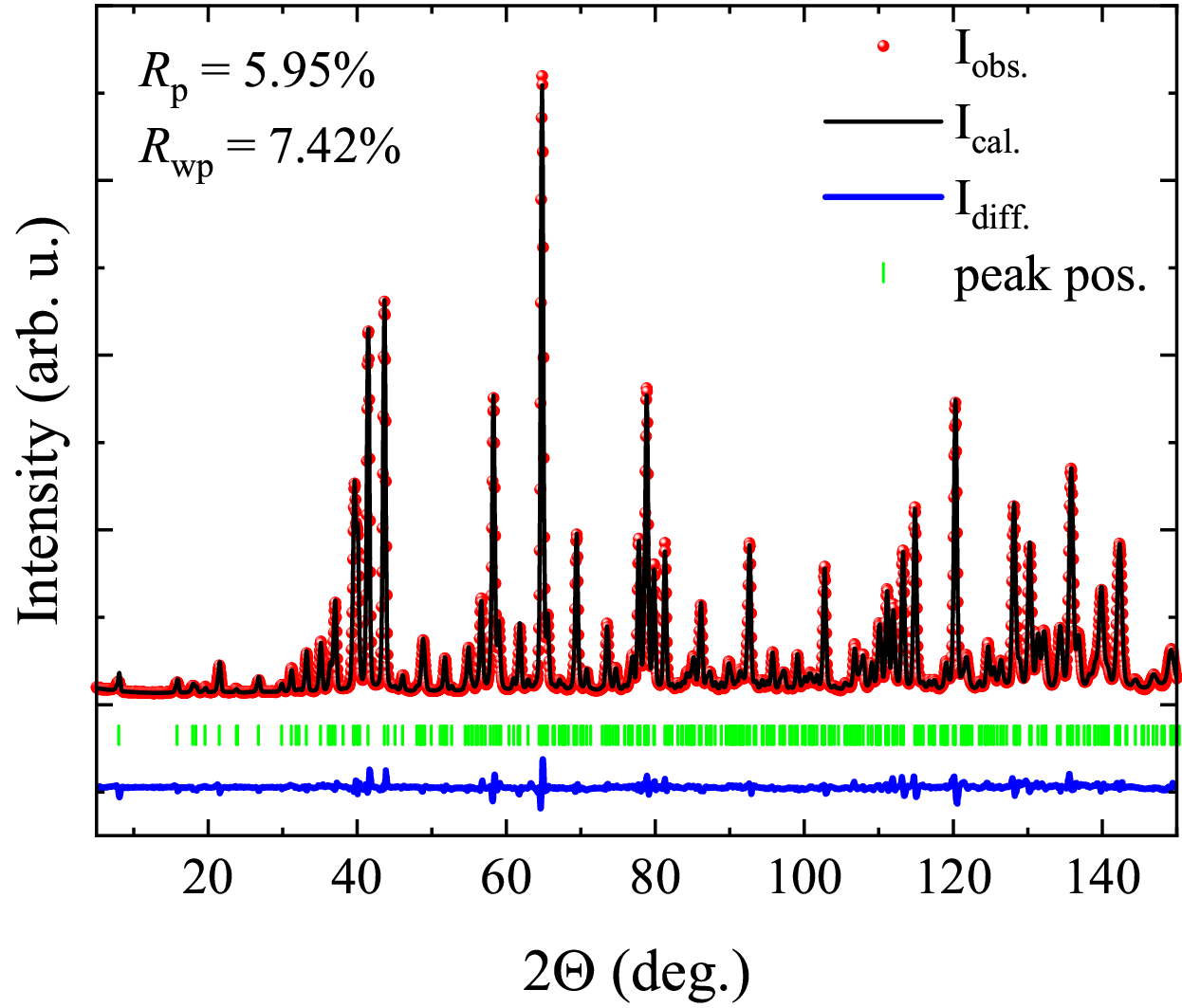}\\
  \caption{Rietveld refinement of the neutron powder diffraction pattern measured at 1.5 K.
}\label{NPD}
\end{figure}

Neutron powder diffraction measurement was performed on the HPRT diffractometer at PSI, Switzerland. The data was collected at 1.5 K with a neutron wavelength of 1.49 $\mathrm{\AA}$. The pattern is shown in Fig. \ref{NPD} with a Rietveld refinement. All the peaks can be indexed with the space group $P6_3/mmc$ (No. 194).

\begin{figure}
  \centering
  \includegraphics[width=1\columnwidth]{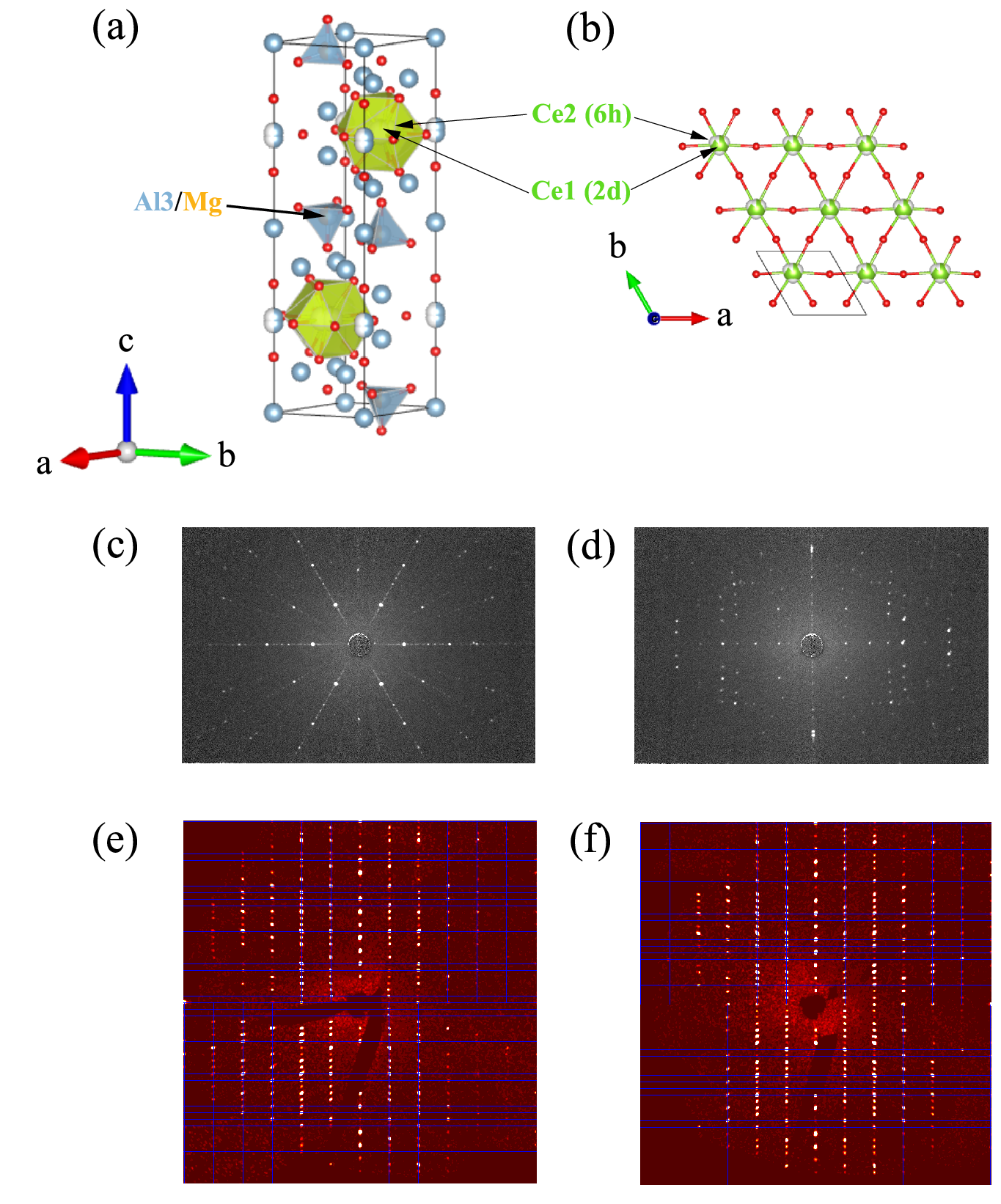}\\
  \caption{Crystal structure of \cmao. (a) The three dimensional view. (b) The top view of the Ce-O layer. (c) Laue patterns with incident X-ray along the $c^{\ast}$ direction and (d) along the $a^{\ast}$ direction. (e,f) Typical single crystal precession images within the 0KL and H0L planes.
}\label{laue}
\end{figure}

More accurate crystal structure is determined by single crystal X-ray diffraction measurement. The experimental conditions and refined crystal structure are listed in Tab. \ref{SXRD}. As was observed for the \pmao\ sample, about $\sim$13\% of the Ce ions are displaced from the 2\textit{d} site to the 6\textit{h} site. The typical crystal structure is shown in Fig. \ref{laue}(a, b). The sharp Laue spots in Fig. \ref{laue}(c, d) indicate a high quality of our crystals.

\begin{table}
\caption{Experimental conditions for the single crystal X-ray diffraction measurements, and the refined crystal structure.\label{SXRD}}
\begin{tabularx}{\linewidth}{X X X X X}
\toprule
\multicolumn{3}{l}{Formula} & \multicolumn{2}{l}{CeMgAl$_{11}$O$_{19}$} \\
\multicolumn{3}{l}{Space group} & \multicolumn{2}{l}{$P6_3/mm\/c$ (No. 194)}\\
\multicolumn{3}{l}{\textit{a, b} ($\textrm{\AA}$)} & \multicolumn{2}{l}{5.59030(10)} \\
\multicolumn{3}{l}{\textit{c} ($\textrm{\AA}$)} & \multicolumn{2}{l}{21.9337(5)} \\
\multicolumn{3}{l}{\textit{V} ($\textrm{\AA}^3$)} & \multicolumn{2}{l}{593.63(2)} \\
\multicolumn{3}{l}{\textit{Z}} & \multicolumn{2}{l}{2} \\
\multicolumn{3}{l}{2$\Theta$ ($^\circ$)} & \multicolumn{2}{l}{3.72 - 102.5} \\
\multicolumn{3}{l}{No. of reflections, $R_\mathrm{int}$} & \multicolumn{2}{l}{20772, 5.36\%}\\
\multicolumn{3}{l}{No. of independent reflections} & \multicolumn{2}{l}{1323} \\
\multicolumn{3}{l}{No. of independent parameters} & \multicolumn{2}{l}{48} \\
\multicolumn{3}{l}{\multirow{3}*{Index ranges}} & -11 $\leq$ \textit{H} $\leq$ 12, \\
\multicolumn{3}{l}{}       & -11 $\leq$ \textit{K} $\leq$ 12,  \\
\multicolumn{3}{l}{}       & -32 $\leq$ \textit{L} $\leq$ 48 \\
\multicolumn{3}{l}{\textit{R}, \textit{wR$_2$}} &\multicolumn{2}{l}{3.12\%, 6.01\%} \\
\multicolumn{3}{l}{Goodness of fit on $F^2$} & \multicolumn{2}{l}{1.36}\\
\multicolumn{3}{l}{Largest difference peak/hole (\textit{e}/$\textrm{\AA}^{3}$)} & \multicolumn{2}{l}{0.91/-1.31}\\
\end{tabularx}

\begin{tabularx}{\linewidth}{X X X X X}
\hline
Atom  & occ. &  x & y & z  \\
Ce1 (2d)  &  0.868(5)     &  0.3333        & 0.6667           & 0.75  \\
Ce2 (6h)  &  0.044(2)   & 0.301(2)     & 0.699(2)      & 0.75 \\
Al1 (2a)  &  1            & 0            & 0             & 0  \\
Al2 (4f)  &  1            & 0.3333       & 0.6667        & 0.18982(3)          \\
Al3 (4f)  & 0.5           & 0.3333       & 0.6667        & 0.47275(3) \\
Mg  (4f)  & 0.5           & 0.3333       & 0.6667        & 0.47275(3) \\
Al4 (12k) & 1             & 0.16745(4)   & 0.33491(7)    & 0.608164(18)        \\
Al5 (4e)  & 0.5           &  0           & 0             & 0.24157(9)   \\
O1  (6h)  & 1             & 0.18093(12)  &   0.3619(2)   &   0.25  \\
O2  (12k) & 1             & 0.15232(9)   &   0.30464(17) &   0.44635(4) \\
O3  (12k) & 1             & 0.50515(16) &   0.49485(8)  &   0.15128(4) \\
O4  (4f)  & 1             & 0.3333       &  0.6667       &   0.55800(7)  \\
O5  (4e)  & 1             & 0            &  0            &   0.34883(7) \\
\end{tabularx}

\begin{tabularx}{\linewidth}{X X X X }
\hline
Atom  & U$_{11}$/U$_{12}$  & U$_{22}$/U$_{13}$  & U$_{33}$/U$_{23}$\\
Ce1   & 0.00794(9)& 0.00794(9)& 0.00509(10) \\
      & 0.00397(5)& 0& 0 \\
Ce2   & 0.069(3)& 0.069(3)& 0.027(3) \\
      & -0.056(4) & 0& 0 \\
Al1   & 0.00392(19)& 0.00392(19)& 0.0040(3) \\
      & 0.00196(10)& 0& 0 \\
Al2   & 0.00424(15)& 0.00424(15)& 0.0035(2) \\
      & 0.00212(7)& 0& 0 \\
Al3   & 0.00374(15)& 0.00374(15)& 0.0044(3) \\
      & 0.00187(8) & 0& 0 \\
Mg    & 0.00374(15)& 0.00374(15)& 0.0044(3) \\
      & 0.00187(8) & 0& 0 \\
Al4   & 0.00403(11)& 0.00400(14)& 0.00449(15) \\
      & 0.00200(7) &-0.00005(5)&-0.00010(9) \\
Al5   & 0.0042(2)& 0.0042(2)& 0.0145(12) \\
      & 0.00210(12) & 0 & 0 \\
O1    & 0.0096(4)& 0.0049(4)& 0.0049(4) \\
      & 0.0024(2)& 0& 0 \\
O2    & 0.0065(2)& 0.0088(3)& 0.0062(3) \\
       & 0.00441(16)& 0.00122(12)& 0.0024(2) \\
O3    & 0.0054(3)& 0.00461(20) & 0.0058(3) \\
      & 0.00271(13)& 0.0014(2)& 0.00069(11) \\
O4    & 0.0046(3)& 0.0046(3)& 0.0073(5) \\
      & 0.00230(14) & 0& 0 \\
O5    & 0.0047(3)& 0.0047(3)& 0.0076(5) \\
     & 0.00236(14) & 0& 0 \\
\toprule
\end{tabularx}
\end{table}

\textbf{2. X-ray Photoelectron Spectroscopy}

\begin{figure}
  \centering
  \includegraphics[width=0.9\columnwidth]{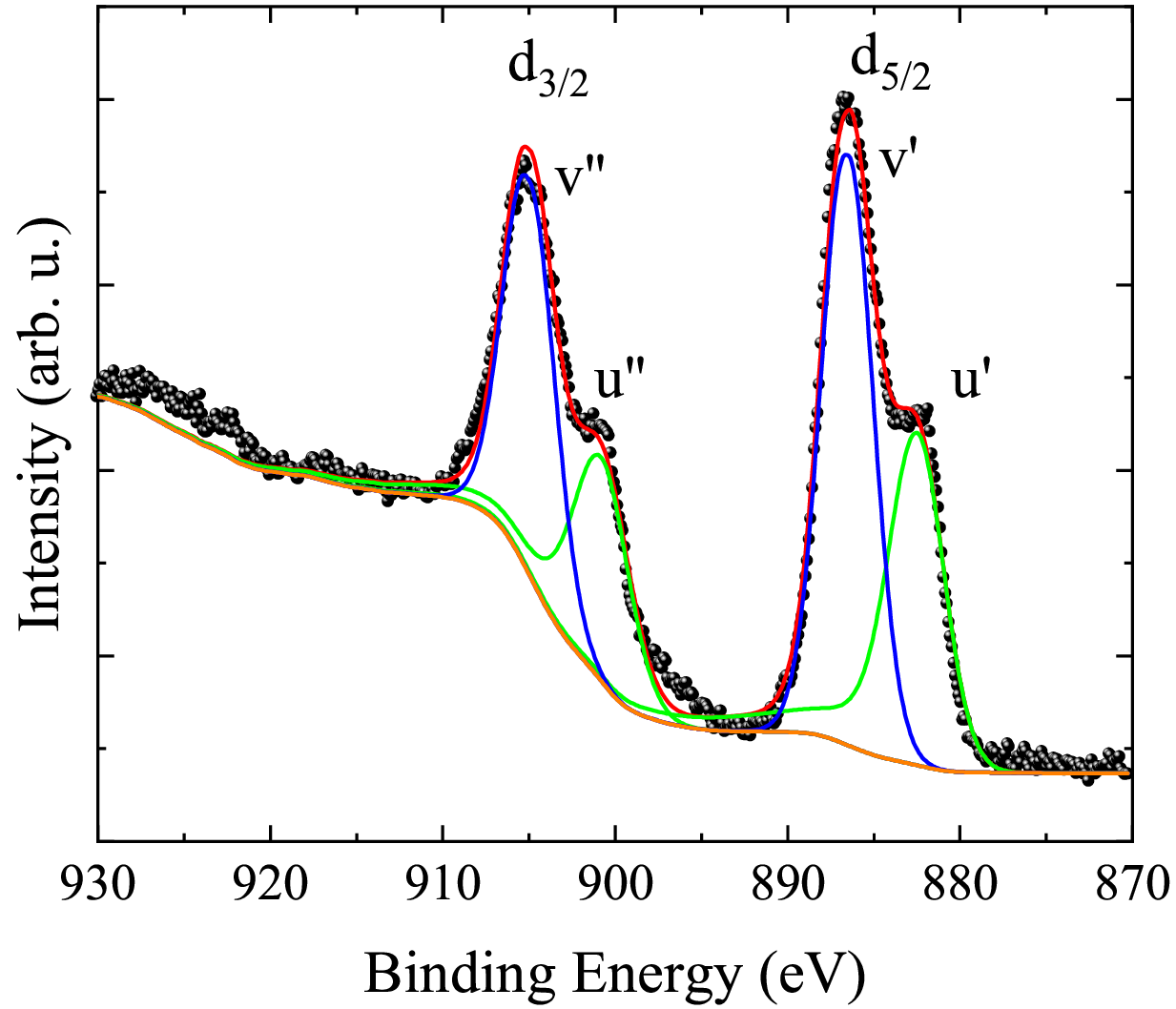}\\
  \caption{Ce 3\textit{d} X-ray photoelectron spectrum for \cmao.
}\label{xps}
\end{figure}

\begin{table}
\caption{Fitting parameters for the XPS data including the binding energy (BE), full width at half maximum (FWHM) and the area below the peak. \label{xps_parameter}}
\begin{ruledtabular}
\begin{tabular}{llll}
    & BE (eV) & FWHM (eV) & Area \\
 u$^\prime$ & 882.51  & 3.77  & 0.74 \\
 u$^{\prime\prime}$ & 900.80  & 3.77  & 0.49 \\
 v$^\prime$ & 886.49  & 3.51  & 1.00 \\
 v$^{\prime\prime}$ & 904.94  & 3.51  & 0.67 \\
\end{tabular}
\end{ruledtabular}
\end{table}

Since Ce ions are prone to forming the Ce$^{4+}$ state in oxides, we have performed X-ray photoelectron spectroscopy (XPS) measurements to investigate the valence state of Ce in our O$_2$ annealed single crystals. The measurements were carried out on an XPS spectrometer (Thermo Fisher ESCALAB 250X) equipped with a monochromated Al $K_{\alpha}$ X-ray source. The spectra were fitted using the Avantage software.

\begin{figure}
  \centering
  \includegraphics[width=1\columnwidth]{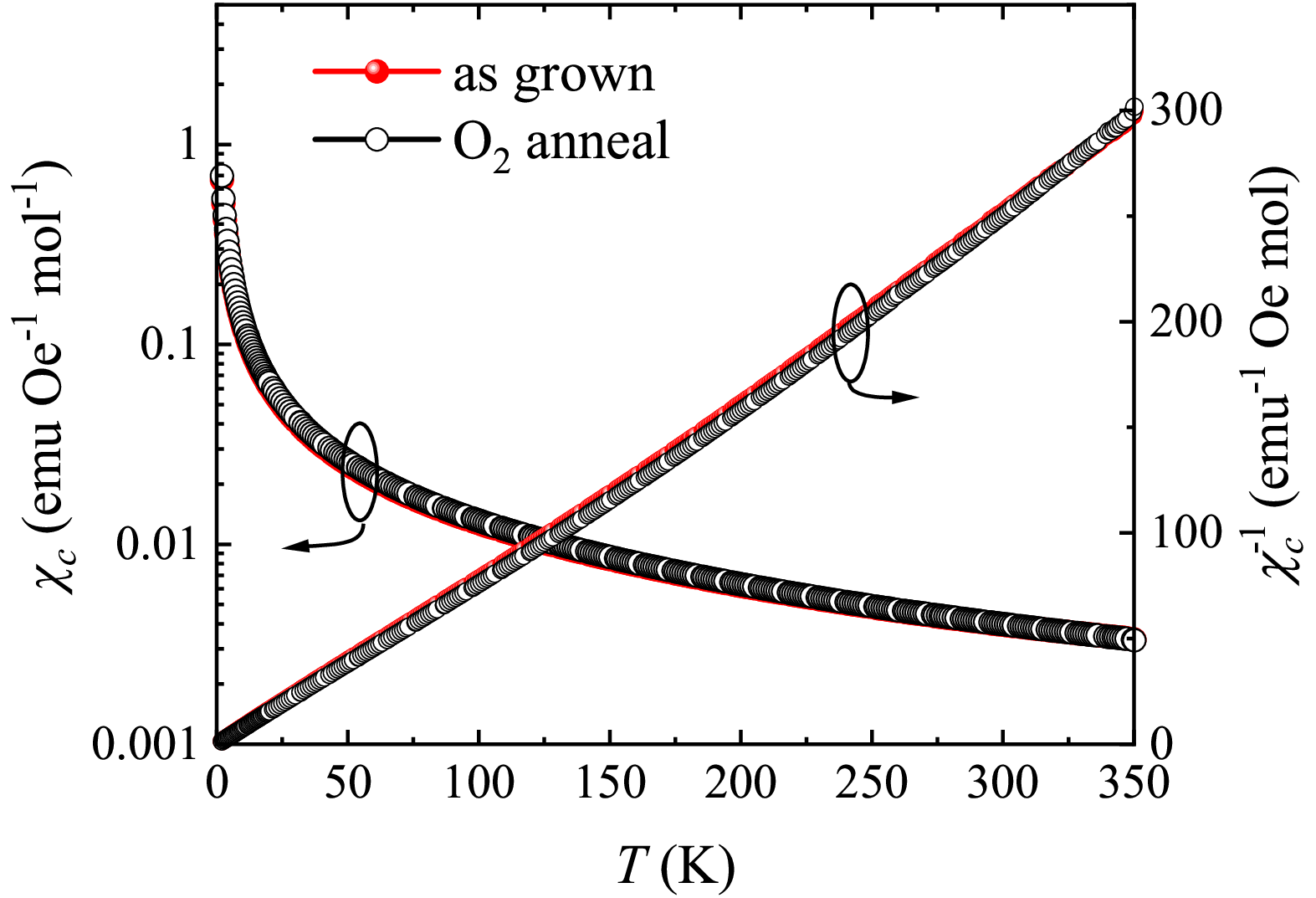}\\
  \caption{Temperature dependence of the magnetic susceptibility $\chi(T)$ (left axis) and inverse magnetic susceptibility $\chi^{-1}(T)$ (right axis) for the as-grown and O$_2$ post-annealed \cmao.}\label{sus2}
\end{figure}

Previous studies have shown that Ce $3d$ XPS can clearly distinguish between Ce$^{4+}$ and Ce$^{3+}$ \cite{beche20083d,teterin1998xps,gurgul2013antimonide}. As shown in Fig. \ref{xps}, the spectrum exhibits two main peaks (v$^\prime$ and v$^{\prime\prime}$), which are due to the spin-orbit spliting of the $3d_{3/2}$ and $3d_{5/2}$ core holes so that the corresponding areas below these peaks should have a ratio of 3:2. Additionally, several satellite peaks (u$^\prime$ and u$^{\prime\prime}$) can be observed due to the multiplet effect. The fitted parameters are shown in Tab. \ref{xps_parameter}. All these features are similar to the results for CePO$_4$ \cite{beche20083d}. Moreover, no marker peak for Ce$^{4+}$ (approximately at 917 eV) was observed. Therefore, our single crystal shows a robust +3 valence state even after O$_2$ annealing. This is contrary to the case of the pyrochlore oxide such as Ce$_2$Zr$_2$O$_7$ where Ce$^{4+}$ can form on the surface of the crystal \cite{Gaudet2019}.
Thus, this compound is very suitable for investigating Ce$^{3+}$-based quantum magnetic states and related magnetic behaviors. Fig. \ref{sus2} compares the magnetic susceptibility measured on the as-grown and O$_2$ annealed samples. No discernible difference can be observed for these two samples.

\textbf{3. CEF fitting}

\begin{figure}
  \centering
  \includegraphics[width=1\columnwidth]{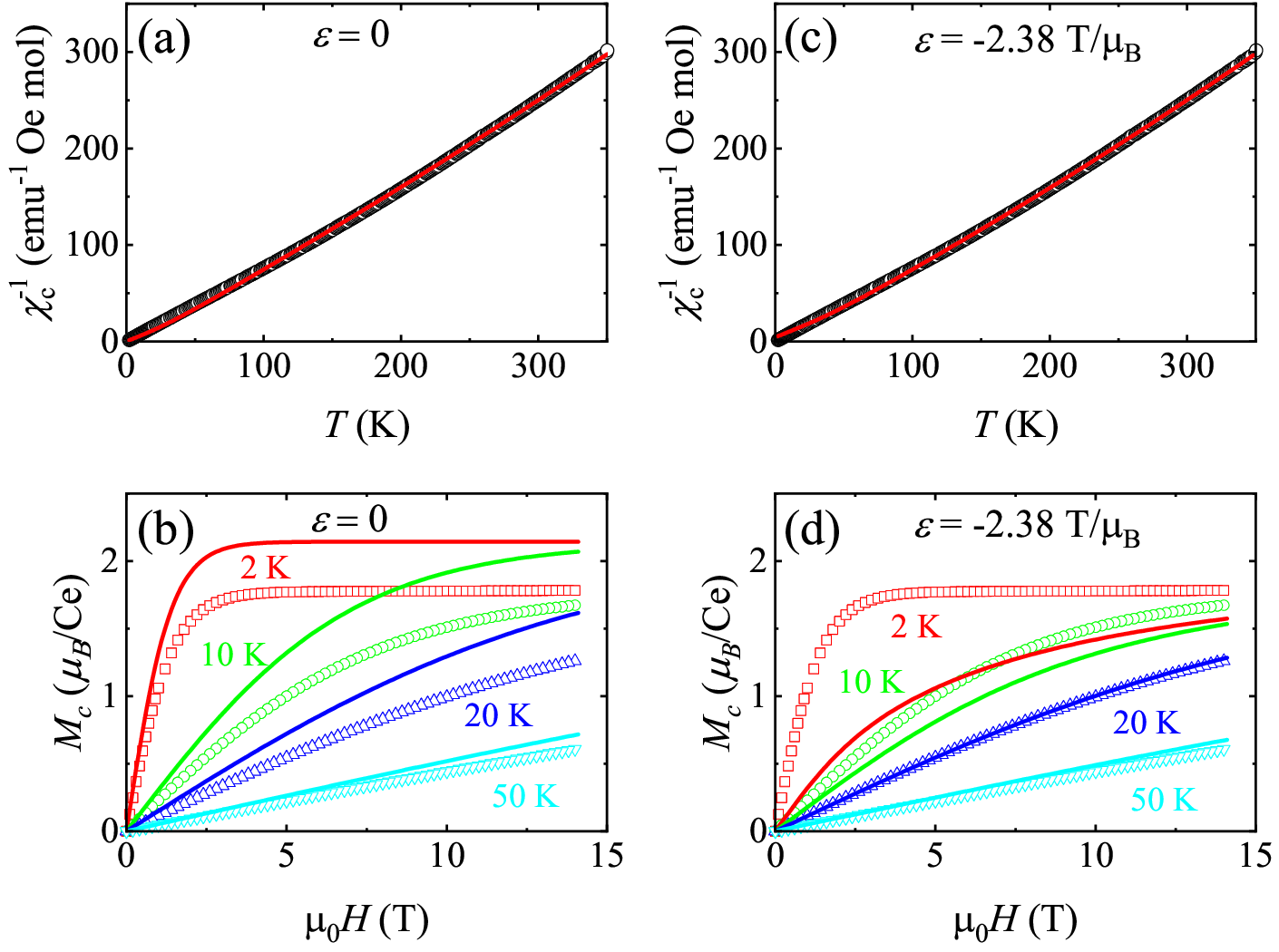}\\
  \caption{CEF simulations based on the \textit{LS} coupled \textit{J} = 5/2 multiplet. In (a) and (b), the mean field parameter $\varepsilon$ is fixed to 0. In (c) and (d), $\varepsilon$ is free to vary during the fitting. The solid curves are calculated from the obtained CEF model.
}\label{J-sim}
\end{figure}

As mentioned in the main text, the CEF models acting on the Hund's rule ground state $^{2}F_{5/2}$ cannot account for the saturation magnetization of \cmao\ at low temperatures. This is further demonstrated in Fig. \ref{J-sim}. The crystal field Hamiltonian is $\mathcal{H}_{CEF} = \sum_{l,m} B_l^mO_l^m$ with only $B_2^0$ and $B_4^0$ are nonzero. We first neglect the mean field effect by setting $\varepsilon$ = 0. The best fit is shown in Fig. \ref{J-sim}(a). The discrepancy between the theory and experiment becomes more evident below $\sim$100 K. This is more apparent when comparing the calculated and experimental magnetization in Fig. \ref{J-sim}(b). After turning on the mean field parameter $\varepsilon$, the saturation magnetization decreases, but the low-field behavior below 10 K deviates significantly from the experimental results; see Fig. \ref{J-sim}(c,d). Therefore, the intermediate coupling scheme as shown in the main text is more suitable for the description of the magnetic behavior of \cmao.

In the main text, we also neglect the impact of the 13\% displaced Ce ions. First, we note that this does not affect the analysis of the saturation magnetization. The XPS results demonstrate that all the Ce ions, either at the 2\textit{d} or the 6\textit{h} site possess a +3 valence state. In the $|J, m_J\rangle$ basis, the saturation magnetization at the 2\textit{d} site is 2.14 $\mu_B$. At the 6\textit{h} site, the CEF ground state could be a linear combination of the $J_z = \pm$1/2, $\pm$3/2 and $\pm$5/2 states. Considering the extreme case of a pure $J_z = \pm$1/2 state, the saturation magnetization at the 6\textit{h} site is 0.43 $\mu_B$. Thus, the total saturation magnetization is 2.14 $\times$ 0.87 + 0.43 $\times$ 0.13 = 1.92 $\mu_B$, which is still much larger than the experimental value (1.77 $\mu_B$). Therefore, even including this displaced Ce ions does not account for the saturation magnetization in the $|J, m_J\rangle$ basis.

\begin{figure}
  \centering
  \includegraphics[width=1\columnwidth]{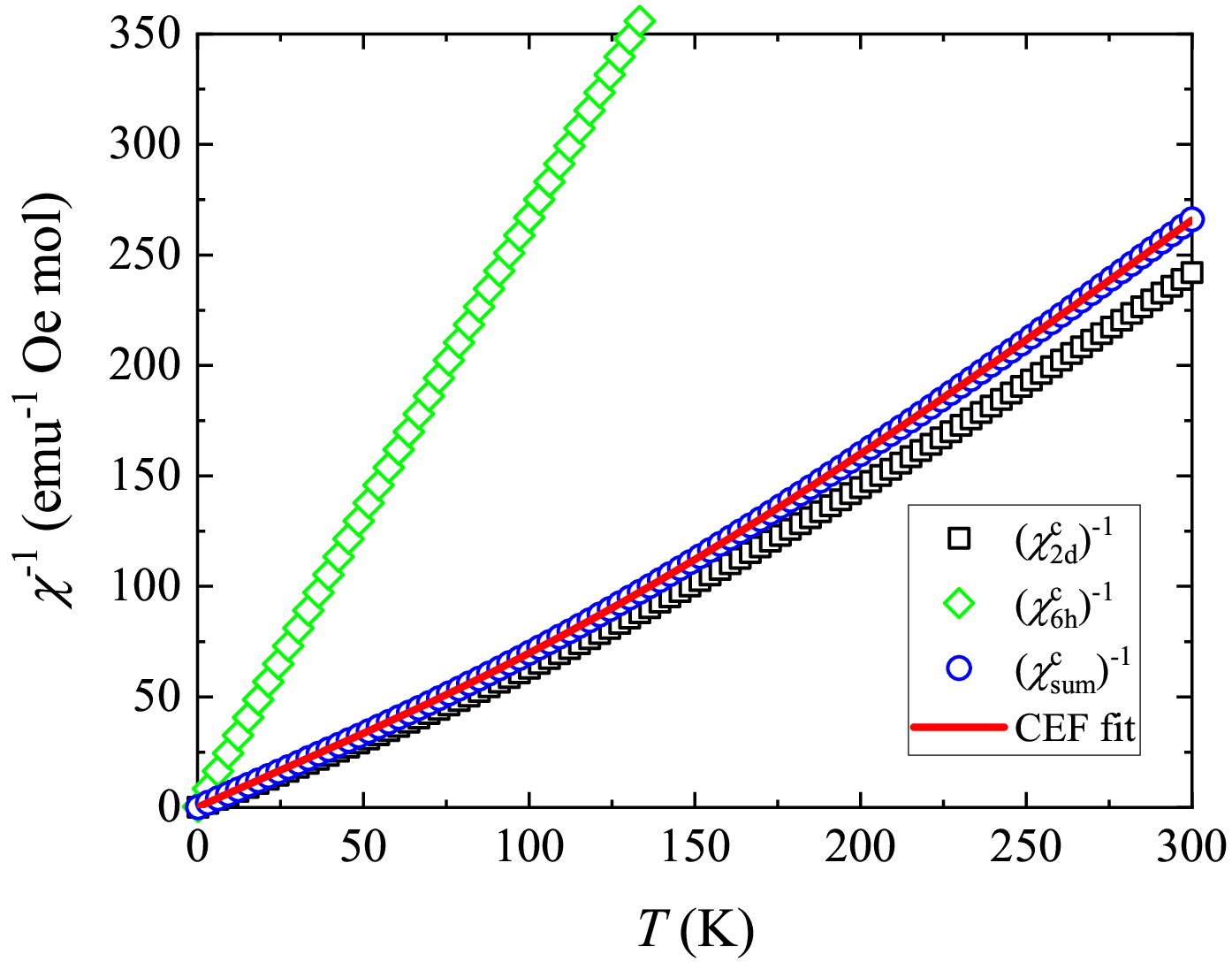}\\
  \caption{Temperature dependence of the inverse magnetic susceptibility calculated from a point charge model for the 2\textit{d} and 6\textit{h} sites. A weighted total susceptibility is also shown together with a CEF fit, see the text for details.}\label{synthetic}
\end{figure}

\begin{figure*}
  \centering
  \includegraphics[width=2\columnwidth]{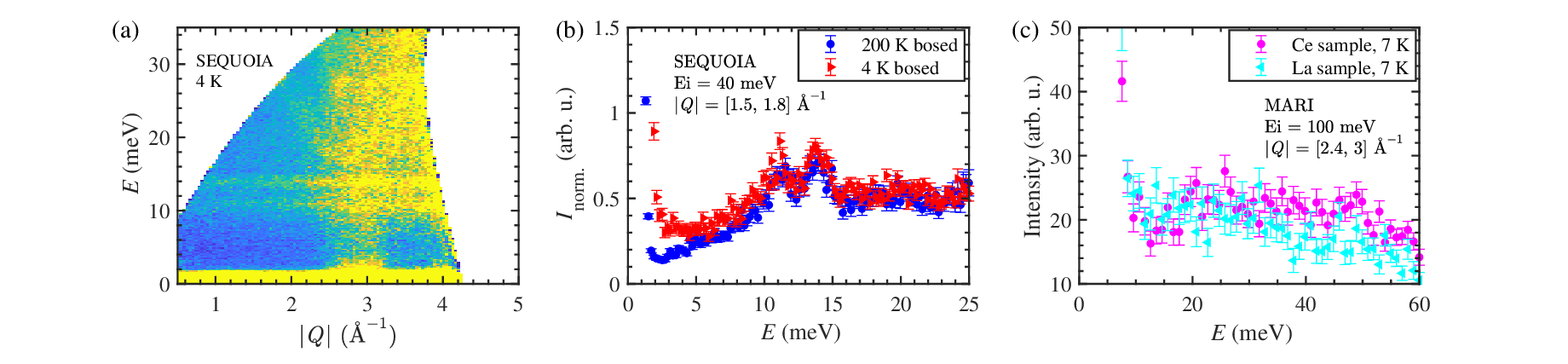}\\
  \caption{(a) Energy (E) - momemtum ($|Q|$) INS map for \cmao\ measured at 4 K. (b) Energy dependence of the $|Q|$-integrated intensities, which are normalized by the Bose factor such that $I_\mathrm{norm.} = S(E)[1-\mathrm{exp}(-E/k_B T)]$. (c) Direct comparison of the $|Q|$-integrated intensities for the \lmao\ and \cmao\ samples measured at 7 K.}\label{INS}
\end{figure*}

Next, we show the influence of the omission of those displaced Ce$^{3+}$ ions in the CEF analysis. For this purpose, we construct the CEF Hamiltonian for the Ce ions at the 2\textit{d} and 6\textit{h} sites based on the point charge model by assuming that the Ce ions are displaced slightly while all the surrounding oxygen positions remain unchanged. From the CEF models, we calculate the magnetic susceptibility, $\chi_{2d}^c$ and $\chi_{6h}^c$, for the Ce ions at the 2\textit{d} and 6\textit{h} sites, respectively. The weighted total susceptibility is obtained as $\chi_\mathrm{total}^c$ = 0.87$\chi_{2d}^c$ + 0.13$\chi_{6h}^c$.
Using this $\chi_\mathrm{total}^c$ as a synthetic experimental data, and following the procedure as described in the main text, we fit the CEF model based on the 2\textit{d} site to it. The eigenenergies and eigenvectors from the best fit are found in Tab. \ref{synthetic_fit}. By comparing this CEF scheme to that from the point charge model, see Tab. \ref{PCM}, one can find that the eigenenergies and engenvectors, especially of the two low-lying states, are not much different. Thus, the omission of those displaced Ce ions in the fitting as described in the main text does not have a significant impact on the CEF scheme.
However, it does have a noticeable effect on the mean-field parameter $\varepsilon$. As shown in Fig. \ref{synthetic}, the synthetic ($\chi_\mathrm{total}^c$)$^{-1}$ is higher than ($\chi_{2d}^c$)$^{-1}$. The best fit yields a negative $\varepsilon$ of -0.065 T/$\mu_B$ (it should be 0 if the displaced Ce$^{3+}$ ions are taken into account in the model). Thus, the omission of those 13\% Ce ions at the 6\textit{d} site may have resulted in an overestimation of the $\varepsilon$ value to the negative side. This explains why a small negative $\varepsilon$ value was observed in our fitting, but a positive Weiss temperature was extracted from the low-\textit{T} CW fit.

\textbf{4. Inelastic neutron scattering}

Inelastic neutron scattering measurements were performed at the SEQUOIA instrument with energies up to 1000 meV on the \cmao\ powders in order to resolve the CEF excitations, which turns out to be very challenging due to the small concentration of the Ce ions. As shown in Fig. \ref{INS}(a), two dispersionless excitations can be observed at $\sim$11 and 14 meV. However, the intensity for these two excitations increases with increasing momentum transfer $|Q|$, indicating that these are more likely to be phonons. In this case, the intensities should be scaled by the Bose factor $1/[1-\mathrm{exp}(-E/k_BT)]$, which is indeed what was observed experimentally; see Fig. \ref{INS}(b). On the other hand, one would expect the CEF excitations following the Boltzmann statistics. Thus, no reliable crystal field levels were observed. Fig. \ref{INS}(c) shows a direct comparison between the La sample and Ce sample measured at MARI, ISIS. No discernible CEF excitation can be observed up to 60 meV.

\begin{table}
\begin{threeparttable}
\caption{Comparison of the lattice parameters, $c/a$ ratio, and unit cell volume of \cmao\ measured experimentally and calculated in this work.\label{lc}}
\begin{tabular}{ccccc}
\toprule
                              & $a$ (\AA)        &  $c$ (\AA)         & $c/a$        & $V_0$ (\AA$^3$) \\
\hline
Expt.                         & 5.59             &  21.93             & 3.92         & 593.87\\
This DFT study                & 5.637            &  22.239            & 3.945        & 611.96 \\
\toprule\\
\end{tabular}
\end{threeparttable}
\end{table}

The phonon excitations are also calculated by first principles using the Vienna \textit{ab initio} simulation package (VASP) \cite{Kresse1993,Kresse1996}, employing the Perdew-Burke-Ernzerhof (PBE) exchange and the projector augmented wave (PAW) method for treatment of the core. The cut-off energy of 550 eV in a plane-wave basis expansion and 8$\times$8$\times$2 $k$-point meshes with their origin at the $\Gamma$ point have been found to provide satisfactory convergence. Due to large conventional unit cell of \cmao, the Gaussian smearing method has been employed in combination with a small smearing width $\sigma$ of 0.05. In order to solve the convergence problems encountered during the selfconsistent calculations, the blocked Davidson algorithm is used, with linear mixing parameter of 0.2 and the cutoff wave vector for Kerker mixing scheme of 0.0001, and the defaulted maximum number of plane-waves is increased. Electronic relaxation is performed until the total energy is converged to 1$\times$10$^{-8}$ eV, and the ionic relaxation is performed until the Hellmann-Feynman forces are less than 0.01 eV/{\AA}.

\begin{figure*}
\centering
\includegraphics[width=15cm]{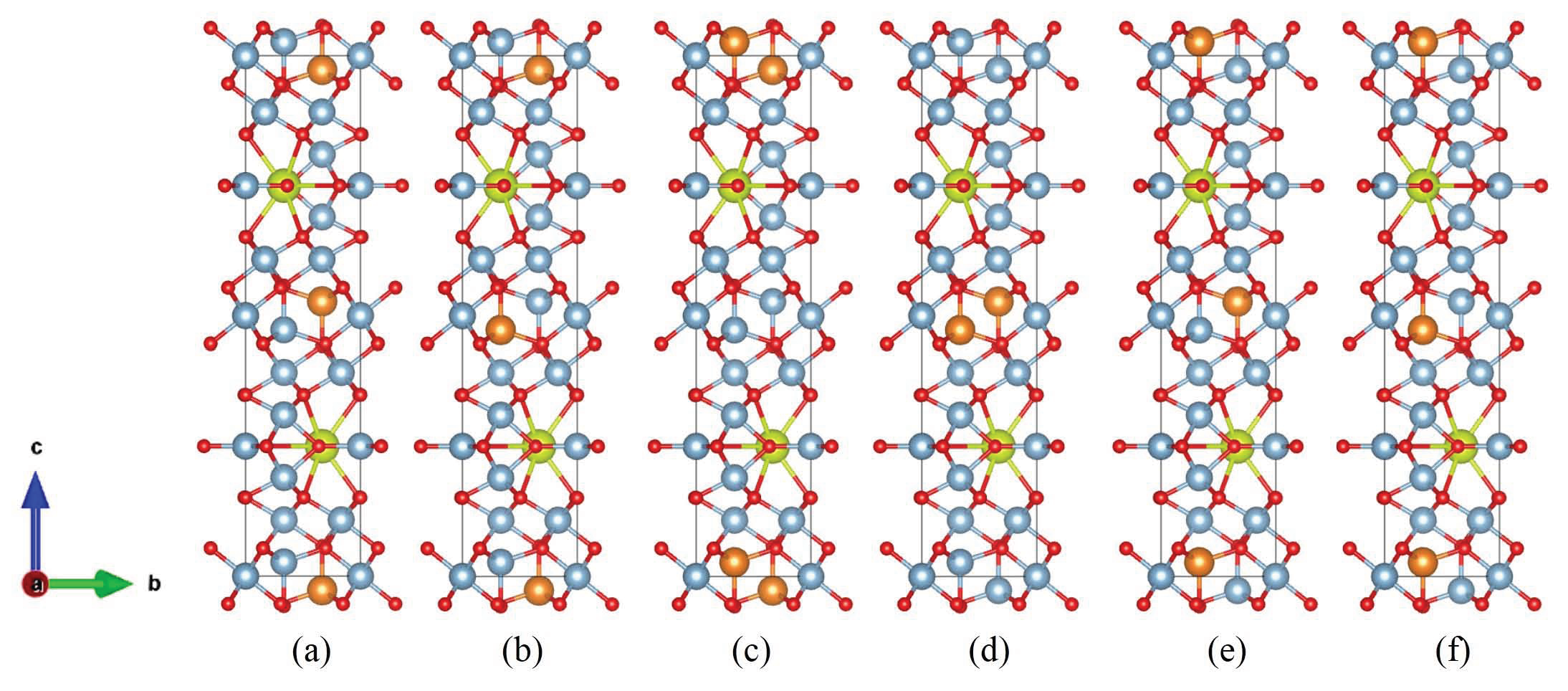}
\hspace{0.5cm}
\caption{Six possible crystal structures of \cmao. Ce, Al, Mg and O atoms are shown in yellow, blue, orange, and red respectively.}\label{structure}
\end{figure*}

The crystal structure of \cmao\ is not perfectly ordered, but there exists certain atomic site disorder. In order to take the atomic site disorder into account, and without increasing the computation time considerably, we have considered 6 possible atomic arrangements for \cmao, as shown in Fig. \ref{structure}. The total energies of all possible crystal structures have been calculated and compared. Our calculations show that the total energies of the relaxed crystal structures $E_a=E_f$, $E_b=E_e$, and $E_c=E_d$. The crystal structure (b) or (e) has the lowest total energy, which is consistent with the refined crystal structure with half occupation of Al3 and Mg at this site. The optimized lattice parameters of $a=5.637$ \AA, and $c=22.239$ \AA, as listed in Tab. \ref{lc}. Comparison with the experimental data $a_0=5.59$ \AA, and $c_0=21.93$ \AA, reveals good agreements. Therefore, the optimized crystal structure (b) has been used for further phonon calculations.

\begin{figure}
  \centering
  \includegraphics[width=1\columnwidth]{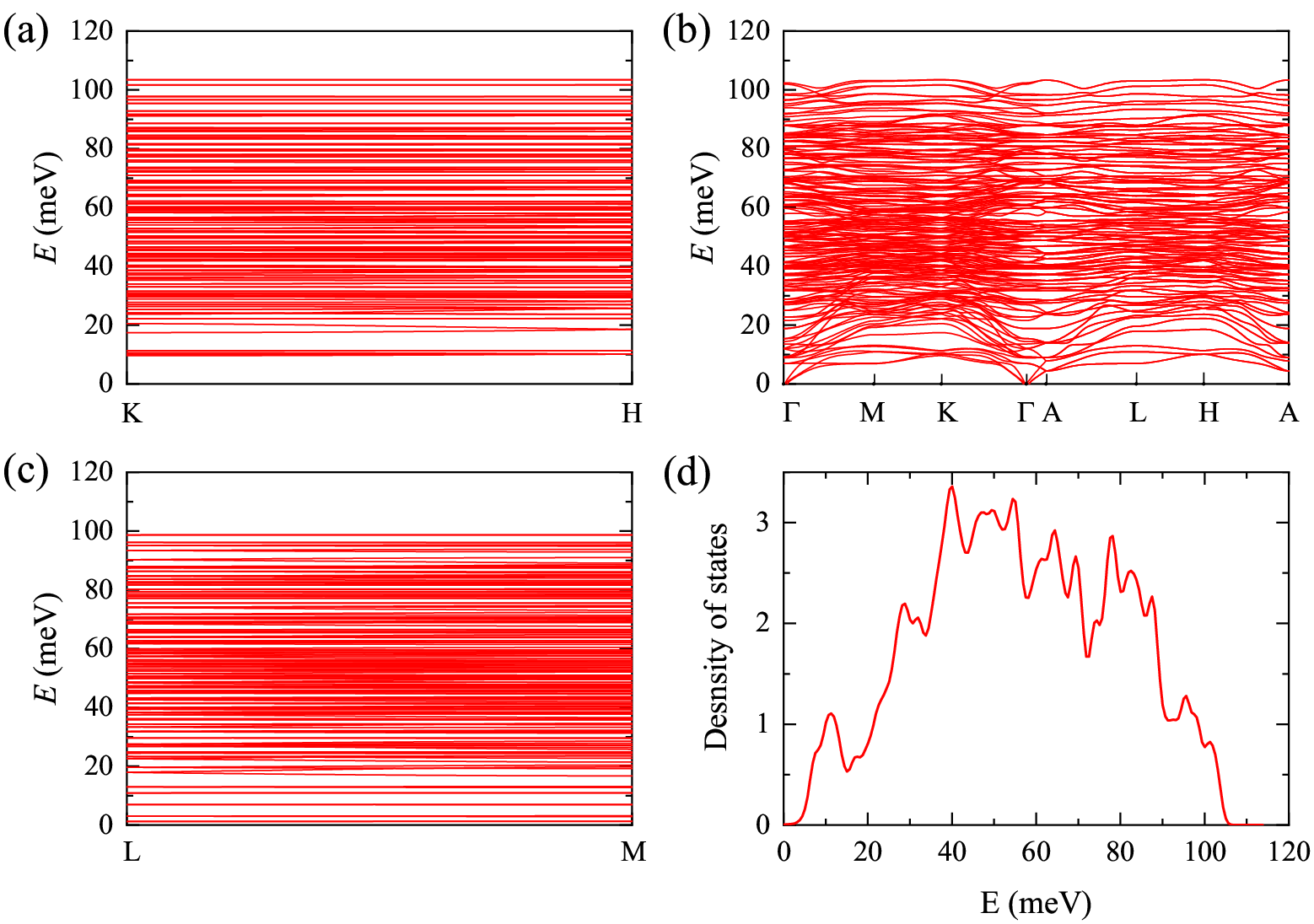}\\
  \caption{(a-c) Calculated phonon dispersions for \cmao\ along high symmetry lines, where $\Gamma=$(0, 0, 0), M=(1/2, 0, 0), K=(1/3, 1/3, 0), A=(0, 0, 1/2), L=(1/2, 0, 1/2), H=(1/3, 1/3, 1/2). (d) Calculated phonon density of states.}\label{phonon}
\end{figure}

Results of our phonon dispersion calculations along the high symmetry directions $\Gamma$-M-K-$\Gamma$-A-L-H-A, K-H, and L-M, in the hexagonal Brillouin zone are shown in Fig. \ref{phonon}(a-c). The phonon density of states (DOS) has been shown in Fig. \ref{phonon}(d). There is no imaginary frequency in all high symmetry directions, which demonstrate that the \cmao\ structure is dynamically stable. As can be seen, substantial DOS is present at $\sim$11 meV, consistent with the INS results.

\textbf{5. Single crystal vs. polycrystal}

\begin{figure}
  \centering
  \includegraphics[width=1\columnwidth]{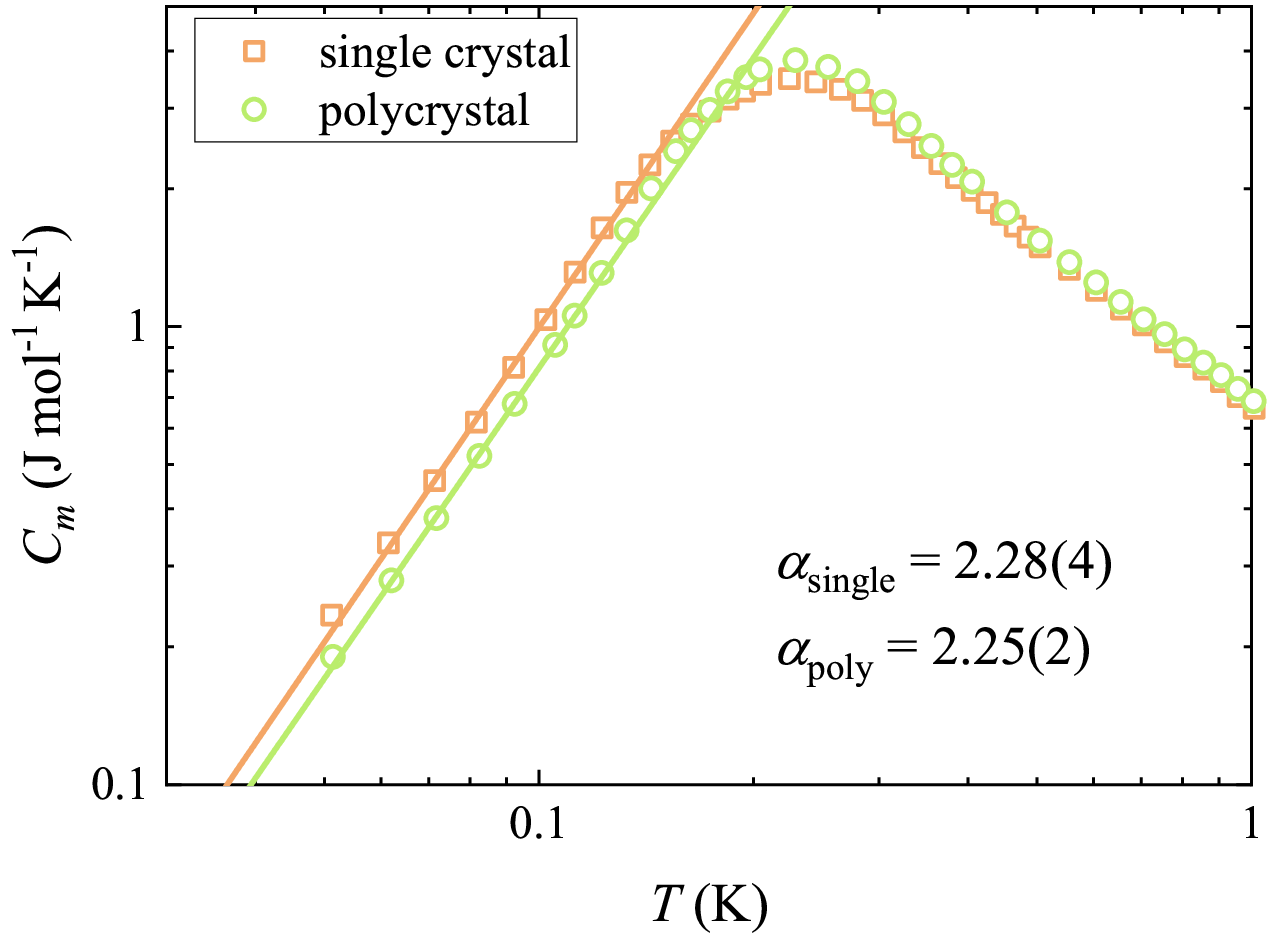}\\
  \caption{Zero-field specific heat comparison between the single crystal and polycrystal. The solid curves are fits to the power law, $C_m = AT^\alpha$.}\label{Cm_comp}
\end{figure}

To clarify potential differences in the sample quality between the single crystal and polycrystal, we present the zero-field specific heat measurements on the two different samples in Fig. \ref{Cm_comp}. As can be seen, both exhibit similar temperature dependence within the whole range. Thus, the difference between the single crystal and polycrystal is negligibly small.


\begin{turnpage}
\begin{table}[t]

\caption{Eigenvectors and Eigenvalues obtained from a CEF fit to the synthetic ($\chi_{total}^c$)$^{-1}$.}
\begin{ruledtabular}
\begin{tabular}{c|cccccccccccccc}
E (meV) &$|-3,-\frac{1}{2}\rangle$ & $|-3,\frac{1}{2}\rangle$ & $|-2,-\frac{1}{2}\rangle$ & $|-2,\frac{1}{2}\rangle$ & $|-1,-\frac{1}{2}\rangle$ & $|-1,\frac{1}{2}\rangle$ & $|0,-\frac{1}{2}\rangle$ & $|0,\frac{1}{2}\rangle$ & $|1,-\frac{1}{2}\rangle$ & $|1,\frac{1}{2}\rangle$ & $|2,-\frac{1}{2}\rangle$ & $|2,\frac{1}{2}\rangle$ & $|3,-\frac{1}{2}\rangle$ & $|3,\frac{1}{2}\rangle$ \tabularnewline
 \hline
0.000 & 0.0 & 0.919 & -0.364 & 0.0 & 0.0 & -0.0 & 0.0 & 0.0 & 0.0 & -0.0 & 0.0 & 0.0 & -0.0 & -0.151 \tabularnewline
0.000 & -0.151 & 0.0 & -0.0 & 0.0 & -0.0 & 0.0 & 0.0 & 0.0 & -0.0 & 0.0 & 0.0 & -0.364 & 0.919 & 0.0 \tabularnewline
21.750 & 0.0 & 0.0 & 0.0 & 0.0 & -0.0 & -0.0 & 0.0 & -0.0 & 0.0 & 0.456 & -0.89 & -0.0 & -0.0 & -0.0 \tabularnewline
21.750 & -0.0 & 0.0 & 0.0 & -0.89 & 0.456 & -0.0 & 0.0 & 0.0 & -0.0 & 0.0 & -0.0 & 0.0 & 0.0 & -0.0 \tabularnewline
44.040 & 0.0 & -0.0 & 0.0 & 0.0 & -0.0 & -0.008 & 0.008 & 0.721 & -0.693 & -0.0 & 0.0 & 0.0 & -0.0 & 0.0 \tabularnewline
44.040 & 0.0 & 0.0 & -0.0 & 0.0 & -0.0 & 0.693 & -0.721 & 0.008 & -0.008 & 0.0 & -0.0 & 0.0 & -0.0 & -0.0 \tabularnewline
265.470 & 0.634 & 0.0 & 0.0 & -0.0 & -0.0 & 0.0 & 0.0 & 0.0 & 0.0 & 0.0 & 0.0 & -0.749 & -0.193 & -0.0 \tabularnewline
265.470 & -0.0 & 0.193 & 0.749 & 0.0 & 0.0 & -0.0 & -0.0 & 0.0 & 0.0 & 0.0 & 0.0 & 0.0 & 0.0 & -0.634 \tabularnewline
300.350 & -0.758 & -0.005 & -0.008 & 0.0 & 0.0 & 0.0 & 0.0 & 0.0 & 0.0 & 0.0 & 0.0 & -0.553 & -0.344 & -0.011 \tabularnewline
300.350 & 0.011 & -0.344 & -0.553 & -0.0 & -0.0 & 0.0 & 0.0 & -0.0 & -0.0 & 0.0 & 0.0 & 0.008 & 0.005 & -0.758 \tabularnewline
314.450 & -0.0 & -0.0 & -0.0 & 0.0 & 0.0 & -0.069 & -0.067 & -0.69 & -0.718 & 0.0 & 0.0 & -0.0 & -0.0 & -0.0 \tabularnewline
314.450 & -0.0 & 0.0 & 0.0 & 0.0 & 0.0 & 0.718 & 0.69 & -0.067 & -0.069 & -0.0 & -0.0 & -0.0 & -0.0 & 0.0 \tabularnewline
325.760 & 0.0 & -0.0 & -0.0 & 0.456 & 0.89 & -0.0 & -0.0 & 0.0 & 0.0 & -0.0 & -0.0 & 0.0 & 0.0 & -0.0 \tabularnewline
325.760 & 0.0 & 0.0 & 0.0 & 0.0 & 0.0 & 0.0 & 0.0 & 0.0 & 0.0 & 0.89 & 0.456 & 0.0 & 0.0 & 0.0 \tabularnewline
\end{tabular}\end{ruledtabular}
\label{synthetic_fit}

\caption{Eigenvectors and eigenvalues for the 2\textit{d} site calculated from the point charge model.\label{PCM}}
\begin{ruledtabular}
\begin{tabular}{c|cccccccccccccc}
E (meV) &$|-3,-\frac{1}{2}\rangle$ & $|-3,\frac{1}{2}\rangle$ & $|-2,-\frac{1}{2}\rangle$ & $|-2,\frac{1}{2}\rangle$ & $|-1,-\frac{1}{2}\rangle$ & $|-1,\frac{1}{2}\rangle$ & $|0,-\frac{1}{2}\rangle$ & $|0,\frac{1}{2}\rangle$ & $|1,-\frac{1}{2}\rangle$ & $|1,\frac{1}{2}\rangle$ & $|2,-\frac{1}{2}\rangle$ & $|2,\frac{1}{2}\rangle$ & $|3,-\frac{1}{2}\rangle$ & $|3,\frac{1}{2}\rangle$ \tabularnewline
 \hline
0.000 & -0.001 & -0.928 & 0.367 & 0.0 & -0.0 & 0.0 & -0.0 & 0.0 & -0.0 & 0.0 & -0.0 & -0.004 & 0.01 & 0.063 \tabularnewline
0.000 & -0.063 & 0.01 & -0.004 & 0.0 & -0.0 & -0.0 & 0.0 & 0.0 & -0.0 & -0.0 & 0.0 & -0.367 & 0.928 & -0.001 \tabularnewline
21.630 & 0.0 & -0.0 & 0.0 & 0.0 & 0.0 & 0.0 & -0.0 & 0.0 & 0.0 & -0.46 & 0.888 & 0.0 & 0.0 & 0.0 \tabularnewline
21.630 & -0.0 & 0.0 & 0.0 & -0.888 & 0.46 & 0.0 & 0.0 & 0.0 & -0.0 & 0.0 & 0.0 & -0.0 & 0.0 & 0.0 \tabularnewline
51.390 & 0.0 & 0.0 & -0.0 & 0.0 & 0.0 & 0.719 & -0.695 & 0.0 & 0.0 & 0.0 & -0.0 & 0.0 & 0.0 & -0.0 \tabularnewline
51.390 & -0.0 & 0.0 & 0.0 & -0.0 & 0.0 & 0.0 & 0.0 & -0.695 & 0.719 & 0.0 & 0.0 & -0.0 & 0.0 & 0.0 \tabularnewline
269.460 & 0.823 & 0.0 & 0.001 & -0.0 & -0.0 & -0.0 & -0.0 & 0.0 & 0.0 & 0.0 & 0.0 & -0.545 & -0.16 & -0.001 \tabularnewline
269.460 & -0.001 & 0.16 & 0.545 & 0.0 & 0.0 & -0.0 & -0.0 & -0.0 & -0.0 & 0.0 & 0.0 & 0.001 & 0.0 & -0.823 \tabularnewline
284.090 & -0.039 & 0.336 & 0.752 & 0.0 & 0.0 & -0.0 & -0.0 & 0.0 & 0.0 & -0.0 & -0.0 & -0.052 & -0.023 & 0.563 \tabularnewline
284.090 & 0.563 & 0.023 & 0.052 & -0.0 & -0.0 & -0.0 & -0.0 & -0.0 & -0.0 & -0.0 & -0.0 & 0.752 & 0.336 & 0.039 \tabularnewline
321.740 & -0.0 & 0.0 & 0.0 & 0.0 & 0.0 & 0.042 & 0.044 & -0.718 & -0.694 & -0.0 & -0.0 & -0.0 & -0.0 & 0.0 \tabularnewline
321.740 & 0.0 & 0.0 & 0.0 & -0.0 & -0.0 & 0.694 & 0.718 & 0.044 & 0.042 & -0.0 & -0.0 & 0.0 & 0.0 & 0.0 \tabularnewline
323.810 & 0.0 & 0.0 & 0.0 & 0.46 & 0.888 & 0.0 & 0.0 & 0.0 & 0.0 & 0.0 & 0.0 & 0.0 & 0.0 & 0.0 \tabularnewline
323.810 & 0.0 & -0.0 & -0.0 & 0.0 & 0.0 & -0.0 & -0.0 & 0.0 & 0.0 & -0.888 & -0.46 & 0.0 & 0.0 & -0.0 \tabularnewline
\end{tabular}
\end{ruledtabular}

\end{table}

\begin{table}
\caption{Eigenvectors and Eigenvalues of the CEF levels in the $|m_L,m_S\rangle$ basis. The spin-orbit coupling strength, $\lambda$, is 78 meV \cite{Abragam}. The extracted CEF parameters from PyCrystalField are $B_2^0$ = -5.2202 meV, $B_4^0$ = 0.0725 meV, $B_6^0$ = 0.0453 meV and $B_6^6$ = -0.1957 meV. The corresponding Wybourne normalised parameters are $B_2^0$, = 234.9 meV, $B_4^0$ = 143.6 meV, $B_6^0$ = -699.1 meV, and $B_6^6$ = 198.9 meV, which can be used for other programs such as SPECTRE.}
\begin{ruledtabular}
\begin{tabular}{c|cccccccccccccc}
E (meV) &$|-3,-\frac{1}{2}\rangle$ & $|-3,\frac{1}{2}\rangle$ & $|-2,-\frac{1}{2}\rangle$ & $|-2,\frac{1}{2}\rangle$ & $|-1,-\frac{1}{2}\rangle$ & $|-1,\frac{1}{2}\rangle$ & $|0,-\frac{1}{2}\rangle$ & $|0,\frac{1}{2}\rangle$ & $|1,-\frac{1}{2}\rangle$ & $|1,\frac{1}{2}\rangle$ & $|2,-\frac{1}{2}\rangle$ & $|2,\frac{1}{2}\rangle$ & $|3,-\frac{1}{2}\rangle$ & $|3,\frac{1}{2}\rangle$ \tabularnewline
 \hline
0.000 & 0.218 & 0.0 & 0.0 & 0.0 & 0.0 & 0.0 & 0.0 & 0.0 & 0.0 & 0.0 & 0.0 & -0.375 & 0.901 & 0.0 \tabularnewline
0.000 & 0.0 & -0.901 & 0.375 & 0.0 & 0.0 & 0.0 & 0.0 & 0.0 & 0.0 & 0.0 & 0.0 & 0.0 & 0.0 & -0.218 \tabularnewline
36.240 & 0.0 & 0.0 & 0.0 & -0.957 & 0.29 & 0.0 & 0.0 & 0.0 & 0.0 & 0.0 & 0.0 & 0.0 & 0.0 & 0.0 \tabularnewline
36.240 & 0.0 & 0.0 & 0.0 & 0.0 & 0.0 & 0.0 & 0.0 & 0.0 & 0.0 & 0.29 & -0.957 & 0.0 & 0.0 & 0.0 \tabularnewline
90.360 & 0.0 & 0.0 & 0.0 & 0.0 & 0.0 & 0.441 & -0.898 & 0.0 & 0.0 & 0.0 & 0.0 & 0.0 & 0.0 & 0.0 \tabularnewline
90.360 & 0.0 & 0.0 & 0.0 & 0.0 & 0.0 & 0.0 & 0.0 & -0.898 & 0.441 & 0.0 & 0.0 & 0.0 & 0.0 & 0.0 \tabularnewline
256.210 & -0.479 & 0.0 & 0.0 & 0.0 & 0.0 & 0.0 & 0.0 & 0.0 & 0.0 & 0.0 & 0.0 & -0.845 & -0.236 & 0.0 \tabularnewline
256.210 & 0.0 & -0.236 & -0.845 & 0.0 & 0.0 & 0.0 & 0.0 & 0.0 & 0.0 & 0.0 & 0.0 & 0.0 & 0.0 & -0.479 \tabularnewline
321.020 & -0.85 & 0.0 & 0.0 & 0.0 & 0.0 & 0.0 & 0.0 & 0.0 & 0.0 & 0.0 & 0.0 & 0.381 & 0.364 & 0.0 \tabularnewline
321.020 & 0.0 & 0.364 & 0.381 & 0.0 & 0.0 & 0.0 & 0.0 & 0.0 & 0.0 & 0.0 & 0.0 & 0.0 & 0.0 & -0.85 \tabularnewline
431.820 & 0.0 & 0.0 & 0.0 & 0.0 & 0.0 & 0.0 & 0.0 & 0.441 & 0.898 & 0.0 & 0.0 & 0.0 & 0.0 & 0.0 \tabularnewline
431.820 & 0.0 & 0.0 & 0.0 & 0.0 & 0.0 & -0.898 & -0.441 & 0.0 & 0.0 & 0.0 & 0.0 & 0.0 & 0.0 & 0.0 \tabularnewline
480.830 & 0.0 & 0.0 & 0.0 & 0.29 & 0.957 & 0.0 & 0.0 & 0.0 & 0.0 & 0.0 & 0.0 & 0.0 & 0.0 & 0.0 \tabularnewline
480.830 & 0.0 & 0.0 & 0.0 & 0.0 & 0.0 & 0.0 & 0.0 & 0.0 & 0.0 & -0.957 & -0.29 & 0.0 & 0.0 & 0.0 \tabularnewline
\end{tabular}\end{ruledtabular}
\label{Eigenvectors}
\end{table}
\end{turnpage}

\end{document}